\documentclass[journal,12pt,twoside]{IEEEtran}
\pdfoutput=1

\usepackage[utf8]{inputenc}

\ifCLASSINFOpdf
  \usepackage[pdftex]{graphicx}
  \graphicspath{/figures/}
\else
\fi

\usepackage{amsmath}
\usepackage[cmintegrals]{newtxmath}
\usepackage[detect-weight=true, binary-units=true]{siunitx} 
\usepackage[xindy,acronyms,nopostdot]{glossaries} 
\usepackage[dvipsnames]{xcolor}

\usepackage{algorithm}
\usepackage{algpseudocode,setspace}

\usepackage{dsfont}					
\usepackage{paralist}			
\usepackage{multirow}	

\ifCLASSOPTIONcompsoc
  \usepackage[caption=false,font=normalsize,labelfont=sf,textfont=sf]{subfig} 
\else
  \usepackage[caption=false,font=footnotesize]{subfig}
\fi

\usepackage{cite}

\usepackage{bm} 

\usepackage{rotating}

\usepackage{multirow}

\usepackage[final]{changes}
\definechangesauthor[color=green]{td}
\definechangesauthor[color=green]{rs}

\usepackage{makecell}

\DeclareSIUnit{\channeluse}{\ensuremath{channel~use}}
\DeclareSIUnit{\dBW}{\ensuremath{dBW}}
\DeclareSIUnit{\dBi}{\ensuremath{dBi}}
\DeclareSIUnit[per-mode=symbol]{\bps}{b\per\second}
\DeclareSIUnit[per-mode=symbol]{\Gbps}{\giga b\per\second}
\DeclareSIUnit{\bit}{b}

\setacronymstyle{long-short}
\newacronym[description={advanced \gls{af}},user1={Advanced Amplify-and-Forward}]{aaf}{AAF}{advanced amplify-and-forward}
\newacronym{af}{AF}{amplify-and-forward}
\newacronym{awgn}{AWGN}{additive white Gaussian noise} 
\newacronym{cnr}{CNR}{carrier to noise power ratio} 
\newacronym{cir}{CIR}{carrier to interference ratio} 
\newacronym{cinr}{CINR}{carrier to interference plus noise power ratio} 
\newacronym{csi}{CSI}{channel state information} 
\newacronym{ctm}{CTM}{channel transfer matrix} 
\newacronym{dpc}{DPC}{dirty paper coding}
\newacronym{dsn}{DSN}{deep space network}
\newacronym{dtp}{DTP}{digital transparent processing}
\newacronym{ecef}{ECEF}{earth centered, earth fixed}
\newacronym{eirp}{EIRP}{effective isotropic radiated power}
\newacronym{em}{EM}{electromagnetic}
\newacronym{fdma}{FDMA}{frequency division multiple access}
\newacronym[shortplural={FSS},firstplural={fixed satellite services (FSS)}]{fss}{FSS}{fixed satellite service}
\newacronym{geo}{GEO}{geostationary earth orbit} 
\newacronym{hpa}{HPA}{high power amplifier}
\newacronym{hts}{HTS}{high throughput satellite}
\newacronym{iid}{i.i.d.}{independent and identically distributed}
\newacronym{kkt}{KKT}{Karush-Kuhn-Tucker}
\newacronym{lms}{LMS}{land mobile satellite} 
\newacronym{los}{LOS}{line-of-sight}
\newacronym{mimo}{MIMO}{multiple-input-multiple-output}
\newacronym{miso}{MISO}{multiple-input-single-output}
\newacronym{mu-mimo}{MU-MIMO}{multiuser MIMO}
\newacronym{mrc}{MRC}{maximum ratio combining} 
\newacronym{mse}{MSE}{mean squared error}
\newacronym[shortplural={MSS},firstplural={mobile satellite services (MSS)}]{mss}{MSS}{mobile satellite service}
\newacronym[user1={Minimum Mean Square Error}]{mmse}{MMSE}{minimum mean square error}
\newacronym[description={naive \gls{af}},user1={Naive Amplify-and-Forward}]{naf}{NAF}{naive amplify-and-forward}
\newacronym{nasa}{NASA}{national aeronautics and space administration}
\newacronym{nlos}{NLOS}{non-line-of-sight} 
\newacronym{obp}{OBP}{on-board processing}
\newacronym{pdf}{PDF}{probability density function} 
\newacronym{psk}{PSK}{phase-shift keying}
\newacronym{pwm}{PWM}{plane wave model} 
\newacronym{qam}{QAM}{quadrature amplitude modulation}
\newacronym[description={quadrature \gls{psk}}]{qpsk}{QPSK}{quadrature phase-shift keying} 
\newacronym[description={amplitude \gls{psk}}]{apsk}{APSK}{amplitude phase-shift keying} 
\newacronym{qos}{QoS}{quality-of-service}
\newacronym{rf}{RF}{radio frequency} 
\newacronym{rx}{Rx}{receiver} 
\newacronym{satcom}{SATCOM}{satellite communications} 
\newacronym{sdma}{SDMA}{space division multiple access} 
\newacronym{siso}{SISO}{single-input-single-output}
\newacronym{snr}{SNR}{signal-to-noise ratio}
\newacronym{svd}{SVD}{singular value decomposition}
\newacronym{swm}{SWM}{spherical wave model} 
\newacronym{tdma}{TDMA}{time division multiple access} 
\newacronym{thp}{THP}{Tomlinson-Harashima precoding} 
\newacronym{tx}{Tx}{transmitter}
\newacronym{uca}{UCA}{uniform circular array} 
\newacronym{uhf}{UHF}{ultra-high frequency}
\newacronym{ula}{ULA}{uniform linear array} 
\newacronym{vod}{VoD}{Video-on-Demand} 
\newacronym{wlog}{w.l.o.g.}{without loss of generality}
\newacronym{zf}{ZF}{zero-forcing}

\newacronym{ut}{UT}{user terminal}
\newacronym{apc}{PAPC}{per-antenna power constraint}
\newacronym{spc}{SPC}{sum power constraint}
\newacronym{mpi}{MPI}{Moore-Penrose-Inverse}
\newacronym{sinr}{SINR}{signal-to-interference-and-noise ratio}
\newacronym{madoc}{MADOC}{Multiple Antenna Downlink Orthogonal Clustering}
\newacronym{cci}{CCI}{co-channel interference}
\newacronym{ffr}{FFR}{full frequency reuse}
\newacronym{fl}{FL}{forward link}
\newacronym{sfpb}{SFPB}{single-feed-per-beam}
\newacronym{wrt}{w.r.t.}{with respect to}
\newacronym{rms}{RMS}{root mean square}
\newacronym{sc-fde}{SC-FDE}{single-carrier frequency domain equalization}
\newacronym{ofdm}{OFDM}{orthogonal frequency division multiplexing}
\newacronym{papr}{PAPR}{peak-to-average power ratio}
\newacronym{fr4}{FR4}{four color frequency reuse}

\newacronym{dvb-s2}{DVB-S2}{Digital Video Broadcasting - Satellite - Second generation} 
\newacronym{dvb-s2x}{DVB-S2X}{DVB-S2 extensions}
\newacronym{vhts}{VHTS}{very high throughput satellite}
\newacronym{tv}{TV}{television}
\newacronym{ml}{ML}{maximum-likelihood}
\newacronym{cazac}{CAZAC}{constant amplitude zero autocorrelation}
\newacronym{esomp}{ESOMP}{earth stations on mobile platform}

\newcounter{MYtempeqncnt}

\newlength\figureheight
\newlength\figurewidth
\newlength\loclen

\begin{document}

\title{MIMO Applications for Multibeam Satellites}

\newcommand{\B}[1]{{\mathbf{#1}}}


\newcommand{\Real}{\Re}
\newcommand{\Imag}{\Im}
\newcommand{\Exp}{\operatorname{E}}
\newcommand{\Det}{\operatorname{det}}
\newcommand{\rank}{\operatorname{rank}}
\newcommand{\diag}{\operatorname{diag}}
\newcommand{\intd}{{\,\operatorname{d}}}
\newcommand{\e}{e}
\newcommand{\Tr}{\operatorname{T}}
\newcommand{\He}{\operatorname{H}}
\newcommand{\mpi}{\operatorname{+}}
\newcommand{\conj}{*}
\newcommand{\Prob}{\mathcal{P}}
\newcommand{\func}{f}
\newcommand{\imj}{j}
\newcommand{\mean}{\operatorname{mean}}
\newcommand{\mymin}{\operatorname{min}}
\newcommand{\mymax}{\operatorname{max}}
\newcommand{\fouriertransform}{\mathcal{F}}
\newcommand{\trace}{\operatorname{tr}}

\newcommand{\myreal}[1]{{\Real\left\{ #1 \right\}}}
\newcommand{\myimag}[1]{{\Imag\left\{ #1 \right\}}}
\newcommand{\myexpect}[1]{{\Exp\left\{ #1 \right\}}}
\newcommand{\mydet}[1]{\Det\left( #1 \right)}
\newcommand{\mylog}[2]{\log_{#1}\left(#2\right)}
\newcommand{\myrank}[1]{\ensuremath{\rank\left\{#1\right\}}}
\newcommand{\mydiag}[1]{\ensuremath{\diag\left\{#1\right\}}}
\newcommand{\myDFT}[2]{\fouriertransform_{#1}\left\{#2\right\}}
\newcommand{\myIDFT}[2]{\fouriertransform_{#1}^{-1}\left\{#2\right\}}
\newcommand{\mymaxvalue}[1]{\mymax\left\{#1\right\}}
\newcommand{\myminvalue}[1]{\mymin\left\{#1\right\}}
\newcommand{\mymeanvalue}[1]{\mean\left\{#1\right\}}
\newcommand{\myabs}[1]{\ensuremath{\left|#1\right|}}
\newcommand{\myarg}[1]{\arg\left\{#1\right\}}
\newcommand{\mynorm}[1]{\left\|#1\right\|}
\newcommand{\mytrace}[1]{\trace\left\{#1\right\}}

\newcommand{\dirac}[1]{\delta\left(#1\right)}

\newcommand{\myprob}[1]{\Prob\left(#1\right)}
\newcommand{\myfuncof}[1]{\func\left(#1\right)}

\newcommand{\mymatrix}[1]{\B{#1}}
\newcommand{\myvector}[1]{\B{#1}}
\newcommand{\matrixentry}[3]{\left[#1\right]_{#2,#3}}

\newcommand{\mybessel}[2]{J_{#1}\left(#2\right)}

\newcommand{\mynull}[1]{\ensuremath{\B{0}_{#1}} }
\newcommand{\myone}[1]{\ensuremath{\B{1}_{#1}} }
\newcommand{\myeye}[1]{\ensuremath{\B{I}_{#1}} }

\newcommand{\mye}[2]{{\e^{#1\imj #2}}}
\newcommand{\myexp}[2]{{\exp\left\{#1\imj #2\right\}}}

\newcommand{\ComplexNormalDistribution}[2]{\mathcal{CN}\left(#1,#2\right)}


\newcommand{{\opt}}{\mathrm{opt}}
\newcommand{\key}{\mathrm{key}}
\renewcommand{\ul}{\mathrm{u}}
\newcommand{{\dl}}{\mathrm{d}}
\newcommand{{\Tx}}{\mathrm{Tx}}
\newcommand{{\Rx}}{\mathrm{Rx}}
\newcommand{{\SL}}{\mathrm{S}}
\newcommand{{\ES}}{\mathrm{E}}
\newcommand{\FE}{\mathrm{FE}}
\newcommand{\maxi}{\mathrm{max}}
\newcommand{\mini}{\mathrm{min}}
\newcommand{\ulacenter}{}

\newcommand{\orbit}{o}
\newcommand{\const}{\text{const.}}


\newcommand{\cazaclength}{L_c}
\newcommand{\cazaccounter}{n}
\newcommand{\cazacinteger}{K_c}
\newcommand{\cazac}{c}

\newcommand{\EIRP}{EIRP}
\newcommand{\GoT}{G/T}
\newcommand{\CoN}{C/N}
\newcommand{\CoNo}{\CoN_0}

\newcommand{\SNR}[2]{\rho_{#1#2}}
\newcommand{\stddev}{\sigma}
\newcommand{\variance}{\stddev^2}
\newcommand{\CNR}{\text{CNR}}
\newcommand{\CNRrxdl}[1]{\text{CNR}_{\dl#1}}
\newcommand{\CINR}{\text{CINR}}
\newcommand{\CNRtxul}{\text{CNR}_{\Tx,\ul}}
\newcommand{\CINRtxul}{\text{CINR}_{\Tx,\ul}}
\newcommand{\CINRrxdl}[1]{\text{CINR}_{\dl#1}}
\newcommand{\CINRtxSisoUl}{\CINRtxul^{\text{S}}}
\newcommand{\CINRZF}{\CINR^{\text{ZF}}}
\newcommand{\CNRrxdlZF}{\CNRrxdl{}^{\text{ZF}}}
\newcommand{\CnrAtBeamCenter}{\text{CNR}^\text{bc}}
\newcommand{\CINRuser}[1]{\CINR_{#1}}

\newcommand{\infobit}{b_l}
\newcommand{\decodedinfobit}{\hat{b}_l}
\newcommand{\BEP}{P_{bit}}

\newcommand{\speedoflight}{c_0}
\newcommand{\wavelength}{\lambda}
\newcommand{\frequency}{f}
\newcommand{\freqinstance}{q}

\newcommand{\carrierfreq}{\frequency_c}
\newcommand{\carrierfreqdl}[1]{\frequency_{#1}^{(\dl)}}
\newcommand{\carrierfrequl}[1]{\frequency_{#1}^{(\ul)}}

\newcommand{\carrierwavelength}{\wavelength_c}
\newcommand{\carrierwavelengthul}{\carrierwavelength^{(\ul)}}
\newcommand{\carrierwavelengthdl}{\carrierwavelength^{(\dl)}}

\newcommand{\NumCarrier}{N_c}

\newcommand{\angularfreq}[1]{\omega_{#1}}
\newcommand{\angularfreqsl}[1]{\angularfreq{#1}^{(\SL)}}

\renewcommand{\time}{t}
\newcommand{\timeref}{\time_0}
\newcommand{\deltatime}{\triangle\time}
\newcommand{\timevariant}{\tau}

\newcommand{\pathdelay}[2]{\timevariant_{#1#2}}
\newcommand{\pathdelayul}[2]{\pathdelay{#1}{#2}^{(\ul)}}
\newcommand{\pathdelaydl}[2]{\pathdelay{#1}{#2}^{(\dl)}}
\newcommand{\pathdelaysl}[2]{\pathdelay{#1}{#2}^{(\SL)}}

\newcommand{\timeinstance}{k}
\newcommand{\period}{T}
\newcommand{\symbolperiod}{\period_s}
\newcommand{\systembandwidth}{B_w}
\newcommand{\userbandwidth}{W_{\dl}}
\newcommand{\samplingperiodtx}{\period_t}
\newcommand{\samplingperiodrx}{\period_r}
\newcommand{\samplingfreqtx}{\frequency_t}
\newcommand{\samplingfreqrx}{\frequency_r}

\newcommand{\TransRate}{\mathcal{R}}
\newcommand{\TransRateUser}[1]{\mathcal{R}_{#1}} 
\newcommand{\MaxTransRate}{\mathcal{R}_\maxi}
\newcommand{\TransRatePerBeam}{\overline{\mathcal{R}}}
\newcommand{\DiversityGain}{\mathfrak{d}}
\newcommand{\MultiplexingGain}{\mathfrak{r}}
\newcommand{\MaxMultiplexingGain}{\MultiplexingGain_\maxi}
\newcommand{\MaxDiversityGain}{\DiversityGain_\maxi}
\newcommand{\OptDiversityGain}{\DiversityGain_\opt}

\newcommand{\DeltaRemainder}{\triangle R}

\newcommand{\eccentricity}[1]{e_{#1}}
\newcommand{\eccentricityx}[1]{\eccentricity{#1}^{x}}
\newcommand{\eccentricityy}[1]{\eccentricity{#1}^{y}}
\newcommand{\eccvector}[1]{\B{e}_{#1}}
\newcommand{\deltaeccvector}[1]{\triangle\eccvector{#1}}
\newcommand{\deltaeccentricityx}[1]{\triangle\eccentricityx{#1}}
\newcommand{\deltaeccentricityy}[1]{\triangle\eccentricityy{#1}}

\newcommand{\inclination}[1]{i_{#1}}
\newcommand{\inclinationx}[1]{\inclination{#1}^{x}}
\newcommand{\inclinationy}[1]{\inclination{#1}^{y}}
\newcommand{\inclvector}[1]{\B{i}_{#1}}
\newcommand{\deltainclvector}[1]{\triangle\inclvector{#1}}
\newcommand{\deltainclinationx}[1]{\triangle\inclinationx{#1}}
\newcommand{\deltainclinationy}[1]{\triangle\inclinationy{#1}}

\newcommand{\semimajoraxis}[1]{a_{#1}}
\newcommand{\deltasemimajoraxis}[1]{\triangle\semimajoraxis{#1}}
\newcommand{\semimajoraxisgeo}{\semimajoraxis{G}}

\newcommand{\londriftrate}[1]{D^{(\SL)}_{#1}}
\newcommand{\deltalondriftrate}[1]{\triangle\londriftrate{#1}}

\newcommand{\semiminoraxis}{b}
\newcommand{\focallength}{c}

\newcommand{\apogeeradius}{\distance_a}
\newcommand{\perigeeradius}{\distance_p}

\newcommand{\argumentperigee}[1]{\omega^{(\SL)}_{#1}}
\newcommand{\RAAN}[1]{\Omega^{(\SL)}_{#1}}
\newcommand{\rightasc}[1]{\alpha^{(\SL)}_{#1}}

\newcommand{\orbitradiusant}[1]{r^{(\SL)}_{#1}}

\newcommand{\trueanomaly}{v_{\orbit}}
\newcommand{\orbitradius}{R_{\orbit}}
\newcommand{\SLvelocity}{V}
\newcommand{\SLvelocitymax}{\SLvelocity_{\text{max}}}
\newcommand{\SLheight}{\height_\SL}
\newcommand{\orbitperiod}{\period_{\orbit}}

\newcommand{\earthrotrate}{\Psi}

\newcommand{\height}{h}
\newcommand{\distance}{r}
\newcommand{\distanceant}[2]{\distance_{#1#2}}
\newcommand{\meandistanceant}{\bar{\distance}}
\newcommand{\antspacing}{d}
\newcommand{\spacingopt}{\antspacing_{\opt}}

\newcommand{\anglesuba}[1]{\alpha_{#1}}
\newcommand{\anglesubadl}[1]{\anglesuba{#1}^{(\dl)}}
\newcommand{\anglesubaul}[1]{\anglesuba{#1}^{(\ul)}}
\newcommand{\anglesubb}[1]{\beta_{#1}}
\newcommand{\anglesubbdl}[1]{\anglesubb{#1}^{(\dl)}}
\newcommand{\anglesubbul}[1]{\anglesubb{#1}^{(\ul)}}
\newcommand{\anglesubc}[1]{\psi_{#1}}
\newcommand{\anglesubcdl}[1]{\anglesubc{#1}^{(\dl)}}
\newcommand{\anglesubcul}[1]{\anglesubc{#1}^{(\ul)}}
\newcommand{\subdelta}[2]{\Delta_{#1#2}}

\newcommand{\anglesubda}{\chi_1}
\newcommand{\anglesubdadl}{\anglesubda^{(\dl)}}
\newcommand{\anglesubdaul}{\anglesubda^{(\ul)}}
\newcommand{\anglesubdb}{\chi_2}
\newcommand{\anglesubdbdl}{\anglesubdb^{(\dl)}}
\newcommand{\anglesubdbul}{\anglesubdb^{(\ul)}}
\newcommand{\anglesubdc}{\chi_3}
\newcommand{\anglesubdcdl}{\anglesubdc^{(\dl)}}
\newcommand{\anglesubdcul}{\anglesubdc^{(\ul)}}
\newcommand{\anglesubdd}{\chi_4}
\newcommand{\anglesubdddl}{\anglesubdd^{(\dl)}}
\newcommand{\anglesubddul}{\anglesubdd^{(\ul)}}

\newcommand{\distanceul}{\distance^{(\ul)}}
\newcommand{\distancedl}{\distance^{(\dl)}}
\newcommand{\distanceulant}[2]{\distanceul_{#1#2}}
\newcommand{\distancedlant}[2]{\distancedl_{#1#2}}

\newcommand{\distancemin}{\distance_{\mini}}
\newcommand{\distancemax}{\distance_{\maxi}}
\newcommand{\distanceulmin}{\distanceul_{\mini}}
\newcommand{\distanceulmax}{\distanceul_{\maxi}}
\newcommand{\distancedlmin}{\distancedl_{\mini}}
\newcommand{\distancedlmax}{\distancedl_{\maxi}}

\newcommand{\deltadistanceul}{\triangle\distanceul{}{}}
\newcommand{\deltadistancedl}{\triangle\distancedl{}{}}
\newcommand{\deltadistance}{\triangle\distanceant{}{}}

\newcommand{\ToA}{ToA}
\newcommand{\ToAul}{\ToA^{(\ul)}}
\newcommand{\ToAdl}{\ToA^{(\dl)}}
\newcommand{\deltaToA}{\triangle\ToA}
\newcommand{\deltaToAul}{\triangle\ToAul}
\newcommand{\deltaToAdl}{\triangle\ToAdl}

\newcommand{\lon}{\theta}
\newcommand{\lonsl}{\lon_{\SL}}
\newcommand{\lones}{\lon_{\ES}}
\newcommand{\lontx}{\lon^{(\Tx)}}
\newcommand{\lonrx}{\lon^{(\Rx)}}

\newcommand{\loninitslant}[1]{\lonsl_{0#1}}
\newcommand{\deltaloninitslant}[1]{\triangle\loninitslant{#1}}

\newcommand{\lat}{\phi}
\newcommand{\latsl}{\lat_{\SL}}
\newcommand{\lates}{\lat_{\ES}}
\newcommand{\lattx}{\lat^{(\Tx)}}
\newcommand{\latrx}{\lat^{(\Rx)}}
\newcommand{\latslant}[1]{\latsl_{#1}}
\newcommand{\lattxant}[1]{\lattx_{#1}}
\newcommand{\latrxant}[1]{\latrx_{#1}}

\newcommand{\deltalatslant}[1]{\triangle\latslant{#1}}

\newcommand{\posvec}{\myvector{a}}
\newcommand{\posvectx}{\posvec^{(\Tx)}}
\newcommand{\posvecrx}{\posvec^{(\Rx)}}
\newcommand{\posvecsl}{\posvec^{(\SL)}}
\newcommand{\posveces}{\posvec_{\ES}}
\newcommand{\posvectxant}[1]{\posvectx_{#1}}
\newcommand{\posvecrxant}[1]{\posvecrx_{#1}}
\newcommand{\posvecslant}[1]{\posvec_{\SL,#1}}
\newcommand{\posvecesant}[1]{\posvec_{\ES,#1}}

\newcommand{\xcoord}{x}
\newcommand{\ycoord}{y}
\newcommand{\zcoord}{z}
\newcommand{\xcoordtx}{\xcoord^{(\Tx)}}
\newcommand{\ycoordtx}{\ycoord^{(\Tx)}}
\newcommand{\zcoordtx}{\zcoord^{(\Tx)}}
\newcommand{\xcoordrx}{\xcoord^{(\Rx)}}
\newcommand{\ycoordrx}{\ycoord^{(\Rx)}}
\newcommand{\zcoordrx}{\zcoord^{(\Rx)}}
\newcommand{\xcoordsl}{\xcoord^{(\SL)}}
\newcommand{\ycoordsl}{\ycoord^{(\SL)}}
\newcommand{\zcoordsl}{\zcoord^{(\SL)}}
\newcommand{\xcoordtxant}[1]{\xcoordtx_{#1}}
\newcommand{\ycoordtxant}[1]{\ycoordtx_{#1}}
\newcommand{\zcoordtxant}[1]{\zcoordtx_{#1}}
\newcommand{\xcoordrxant}[1]{\xcoordrx_{#1}}
\newcommand{\ycoordrxant}[1]{\ycoordrx_{#1}}
\newcommand{\zcoordrxant}[1]{\zcoordrx_{#1}}
\newcommand{\xcoordslant}[1]{\xcoordsl_{#1}}
\newcommand{\ycoordslant}[1]{\ycoordsl_{#1}}
\newcommand{\zcoordslant}[1]{\zcoordsl_{#1}}

\newcommand{\antspacinges}{\antspacing_{\ES}}
\newcommand{\antspacingtx}{\antspacing^{(\Tx)}}
\newcommand{\antspacingrx}{\antspacing^{(\Rx)}}
\newcommand{\antspacingsl}{\antspacing_{\SL}}
\newcommand{\antspacingtxopt}{\antspacingtx_{\opt}}
\newcommand{\antspacingrxopt}{\antspacingrx_{\opt}}
\newcommand{\antspacingslopt}{\antspacingsl_{\opt}}
\newcommand{\antspacingopt}{\antspacing_{\opt}}

\newcommand{\antspacingtxant}[1]{\antspacingtx_{#1}}
\newcommand{\antspacingrxant}[1]{\antspacingrx_{#1}}
\newcommand{\antspacingesant}[1]{\antspacing_{\ES,#1}}
\newcommand{\antspacingslant}[1]{\antspacing_{\SL,#1}}

\newcommand{\spacingorbit}{\triangle\lonsl}
\newcommand{\ulaorient}{\delta}
\newcommand{\ulaorienttx}{\ulaorient^{(\Tx)}}
\newcommand{\ulaorientrx}{\ulaorient^{(\Rx)}}
\newcommand{\ulaorientsl}{\ulaorient^{(\SL)}}
\newcommand{\ulaorientes}{\ulaorient_{\ES}}
\newcommand{\ulaorienttxopt}{\ulaorienttx_{\opt}}
\newcommand{\ulaorientrxopt}{\ulaorientrx_{\opt}}
\newcommand{\ulaorientslopt}{\ulaorientsl_{\opt}}

\newcommand{\offaxisangle}[2]{\vartheta_{#1#2}}
\newcommand{\offaxisHPB}{\offaxisangle{3dB}{}}
\newcommand{\diameter}{D}

\newcommand{\gravconst}{G}
\newcommand{\earthmass}{M_{\oplus}}
\newcommand{\earthradius}{R_{\oplus}}
\newcommand{\productGM}{GM_{\oplus}}
\newcommand{\siderealday}{T_{\oplus}}

\newcommand{\AntNumTx}{N}
\newcommand{\AntNumRx}{M}
\newcommand{\AntNumSL}{Z}
\newcommand{\AntNumSLtx}{\AntNumSL_{\mathrm{t}}}
\newcommand{\AntNumSLrx}{\AntNumSL_{\mathrm{r}}}
\newcommand{\countNumTx}{n}
\newcommand{\countNumRx}{m}
\newcommand{\countNumSL}{z}

\newcommand{\counterk}{k}
\newcommand{\counterl}{l}
\newcommand{\countert}{t}
\newcommand{\counteru}{u}

\newcommand{\vfactor}[1]{v_{#1}}

\newcommand{\amplgain}{g}
\newcommand{\amplgainant}[1]{\amplgain_{#1}}
\newcommand{\amplgainTxFE}{\amplgain^{(\Tx)}_{\FE}}
\newcommand{\amplgainsl}[1]{a_{\SL}^{#1}}

\newcommand{\phaseshift}[2]{\upsilon_{#1#2}}
\newcommand{\phaseshiftsl}[2]{\phaseshift{#1}{#2}^{(\SL)}}

\newcommand{\phaseshiftest}[2]{\hat{\vartheta}_{#1#2}}

\newcommand{\CTF}{{h}}
\newcommand{\CTFtot}{{c}}
\newcommand{\CTFul}{\CTF^{(\ul)}}
\newcommand{\CTFdl}{\CTF^{(\dl)}}
\newcommand{\CTFsl}{\CTF^{(\SL)}}

\newcommand{\CTMentry}[2]{\CTF_{#1#2}}
\newcommand{\CTMentryul}[2]{\CTMentry{#1}{#2}^{(\ul)}}
\newcommand{\CTMentrydl}[2]{\CTMentry{#1}{#2}^{(\dl)}}

\newcommand{\CTMentryest}[2]{\hat{\CTF}_{#1#2}}

\newcommand{\CTM}{\mymatrix{H}}
\newcommand{\CTMul}[1]{\CTM_{\ul,#1}}
\newcommand{\CTMdl}{\CTM_\dl}
\newcommand{\CTMsl}[1]{\CTMslblk_{#1}}
\newcommand{\CTMblk}{\CTM_{\ul}}
\newcommand{\CTMslblk}{\mymatrix{F}}
\newcommand{\CTMdlmod}{\tilde{\CTM}_{\dl}}

\newcommand{\CTMvecul}[1]{\mymatrix{\CTF}_{\ul,#1}}

\newcommand{\countf}{l}
\newcommand{\counti}{i}

\newcommand{\CTMest}{\hat{\CTM}}
\newcommand{\CTMestnorm}{\CTMest_{\text{norm}}}
\newcommand{\CTMnorm}{\CTM_{\text{norm}}}
\newcommand{\CTMfro}{\CTM_{\text{fro}}}
\newcommand{\CTMestfro}{\CTMest_{\text{fro}}}

\newcommand{\CTMtilde}{\tilde{\CTM}}
\newcommand{\CTMultilde}[1]{\tilde{\CTM}_{\ul,#1}}
\newcommand{\CTMblktilde}{\tilde{\CTM}_{\ul}}
\newcommand{\CTMdltilde}{\tilde{\CTM}_{\dl}}

\newcommand{\CTMentrytilde}[2]{\tilde{\CTF}_{#1#2}}
\newcommand{\CTMentryultilde}[2]{\tilde{\CTF}^{(\ul)}_{#1#2}}
\newcommand{\CTMentrydltilde}[2]{\tilde{\CTF}^{(\dl)}_{#1#2}}
\newcommand{\CTMentrytot}[2]{\CTFtot_{#1#2}}

\newcommand{\FSL}[2]{a_{#1#2}}
\newcommand{\FSLdl}[1]{\FSL{\dl}{#1}}
\newcommand{\FSLest}[2]{\hat{\FSL{}{}}_{#1#2}}

\newcommand{\AtmImp}[1]{\varsigma_{#1}}
\newcommand{\AtmAtt}[1]{A_{#1}}
\newcommand{\AtmAttDiff}{D_{\AtmAtt{}}}
\newcommand{\AtmPhase}[1]{\xi_{#1}}
\newcommand{\AtmImpRatio}{R_{\myabs{\AtmImp{}}}}

\newcommand{\channelphase}{\varphi}
\newcommand{\channelphaseul}{\channelphase_{\ul}}
\newcommand{\channelphasedl}{\channelphase_{\dl}}

\newcommand{\CIR}[2]{h_{#1#2}}
\newcommand{\CIRul}[2]{\CIR{#1}{#2}^{(\ul)}}
\newcommand{\CIRdl}[2]{\CIR{#1}{#2}^{(\dl)}}
\newcommand{\CIRsl}[2]{\CIR{#1}{#2}^{(\SL)}}

\newcommand{\CIRvec}[2]{\mymatrix{h}_{#1#2}}
\newcommand{\CIRvecul}[2]{\CIRvec{#1}{#2}^{(\ul)}}
\newcommand{\CIRvecdl}[2]{\CIRvec{#1}{#2}^{(\dl)}}

\newcommand{\CIRvecest}[2]{\hat{\mymatrix{h}}_{#1#2}}

\newcommand{\CIMentry}[2]{\CIR_{#1#2}}
\newcommand{\CIMentryul}[2]{\CIMentry{#1}{#2}^{(\ul)}}
\newcommand{\CIMentrydl}[2]{\CIMentry{#1}{#2}^{(\dl)}}

\newcommand{\CIM}{\mymatrix{H}}
\newcommand{\CIMul}{\CIM_\ul}
\newcommand{\CIMdl}{\CIM_\dl}
\newcommand{\CIMsl}{\CIM_\SL}

\newcommand{\CIRtap}{l}
\newcommand{\CIRtapul}{\CIRtap_{\ul}}
\newcommand{\CIRtapdl}{\CIRtap_{\dl}}
\newcommand{\CIRtapmax}{L}
\newcommand{\CIRtapmaxul}{\CIRtapmax_{\ul}}
\newcommand{\CIRtapmaxdl}{\CIRtapmax_{\dl}}

\newcommand{\CTMprod}{\mymatrix{V}}
\newcommand{\CTMprodul}{\CTMprod_{\ul}}
\newcommand{\CTMproddl}{\CTMprod_{\dl}}

\newcommand{\Xmatrix}[1]{\mymatrix{X}_{#1}}
\newcommand{\Smatrix}{\mymatrix{S}}
\newcommand{\Umatrix}{\mymatrix{U}}
\newcommand{\Gammamatrix}{\mymatrix{\Gamma}}
\newcommand{\Umatrixul}{\Umatrix_{\ul}}
\newcommand{\Umatrixdl}{\Umatrix_{\dl}}
\newcommand{\Gammamatrixul}{\Gammamatrix_{\ul}}
\newcommand{\Gammamatrixdl}{\Gammamatrix_{\dl}}
\newcommand{\matrixV}{\mymatrix{V}}
\newcommand{\Dmatrix}{\mymatrix{D}}

\newcommand{\Ematrix}{\mymatrix{E}}
\newcommand{\Rmatrix}{\mymatrix{R}}

\newcommand{\eigenvalue}[1]{\gamma_{#1}}
\newcommand{\eigenvalueul}[1]{\eigenvalue{#1}^{(\ul)}}
\newcommand{\eigenvaluedl}[1]{\eigenvalue{#1}^{(\dl)}}

\newcommand{\maxNumAnt}{V}
\newcommand{\minNumAnt}{U}

\newcommand{\noise}{\eta}
\newcommand{\noiseul}{\noise^{(\ul)}}
\newcommand{\noisedl}{\noise^{(\dl)}}

\newcommand{\noisevec}[1]{\bm{\noise}_{#1}}
\newcommand{\noisevecul}{\noisevec{\ul}}
\newcommand{\noisevecdl}{\noisevec{\dl}}
\newcommand{\noisevecentry}[1]{\noise_{#1}}
\newcommand{\noisevecentryul}[1]{\noise_{\ul,#1}}
\newcommand{\noisevecentrydl}[1]{\noise_{\dl,#1}}
\newcommand{\noiseveceq}{\bar{\noisevec{}}}

\newcommand{\noisepow}[1]{\variance_{\noisevec{#1}}}
\newcommand{\noisepowul}{\variance_{\noisevecul}}
\newcommand{\noisepowdl}[1]{\variance_{\noisevecdl#1}}
\newcommand{\noisepoweq}[1]{\variance_{\noiseveceq#1}}

\newcommand{\noisestddev}{\stddev_{\noisevec}}
\newcommand{\noisestddevul}{\stddev_{\noisevecul}}
\newcommand{\noisestddevdl}{\stddev_{\noisevecdl}}


\newcommand{\noisevecfreq}{\noisevec_{\B{f}}}
\newcommand{\noisevecfrequl}{\noisevec_{\B{f}\ul}}
\newcommand{\noisevecfreqdl}{\noisevec_{\B{f}\dl}}

\newcommand{\constellation}{\mathds{A}}

\newcommand{\symboltx}{s}
\newcommand{\symboltxvec}{\myvector{\symboltx}}
\newcommand{\symboltxvecentry}[1]{\symboltx_{#1}}

\newcommand{\signaltx}{x}
\newcommand{\signaltxvec}[1]{\myvector{\signaltx}_{#1}}
\newcommand{\signaltxvecentry}[1]{\signaltx_{#1}}

\newcommand{\signaltxest}{\hat{\signaltx}}
\newcommand{\signaltxestvec}[1]{\myvector{\signaltxest}_{#1}}
\newcommand{\signaltxestvecentry}[1]{\signaltxest_{#1}}

\newcommand{\signaltxul}{\signaltx^{(\ul)}}
\newcommand{\signaltxdl}{\signaltx^{(\dl)}}
\newcommand{\signaltxvecul}{\signaltxvec{\ul}}
\newcommand{\signaltxvecdl}{\signaltxvec{\dl}}
\newcommand{\signaltxveculentry}[1]{\signaltx_{\ul,#1}}
\newcommand{\signaltxvecdlentry}[1]{\signaltx_{\dl,#1}}

\newcommand{\signaltxpow}[1]{\variance_{\signaltxvec{#1}}}
\newcommand{\signaltxpowant}[1]{\variance_{\signaltxvecentry{#1}}}
\newcommand{\signaltxpowul}{\variance_{\signaltxvecul}}
\newcommand{\signaltxpowantul}[1]{\variance_{\signaltxveculentry{#1}}}
\newcommand{\signaltxpowdl}{\variance_{\signaltxvecdl}}
\newcommand{\signaltxpowantdl}[1]{\variance_{\signaltxvecdlentry{#1}}}

\newcommand{\signaltxvecfreq}{\signaltxvec_{\B{f}}}

\newcommand{\CovMat}{\mymatrix{R}}
\newcommand{\CovMattxsig}{\CovMat_{\signaltxvec{}}}
\newcommand{\CovMattxsigul}{\CovMat_{\signaltxvecul}}
\newcommand{\CovMattxsigdl}{\CovMat_{\signaltxvecdl}}

\newcommand{\CovMatnoise}{\CovMat_{\noisevec{}}}
\newcommand{\CovMatnoiseul}{\CovMat_{\noisevecul}}
\newcommand{\CovMatnoisedl}{\CovMat_{\noisevecdl}}
\newcommand{\CovMatnoisetot}{\CovMat_{\noiseveceq}}

\newcommand{\symbolrx}{\hat{\symboltx}}
\newcommand{\symbolrxvec}{\hat{\symboltxvec}}
\newcommand{\symbolrxvecentry}[1]{\symbolrx_{#1}}

\newcommand{\signalrx}{y}
\newcommand{\signalrxvec}{\mymatrix{\signalrx}}
\newcommand{\signalrxvecul}{\myvector{\signalrx}_{\ul}}
\newcommand{\signalrxvecdl}{\myvector{\signalrx}_{\dl}}
\newcommand{\signalrxvecentry}[1]{\signalrx_{#1}}
\newcommand{\signalrxveculentry}[1]{\signalrx_{\ul,#1}}
\newcommand{\signalrxvecdlentry}[1]{\signalrx_{\dl,#1}}

\newcommand{\signalrelay}{z}
\newcommand{\signalrelayvec}{\myvector{\signalrelay}}
\newcommand{\signalrelayvecentry}[1]{\signalrelay_{#1}}

\newcommand{\signalrxvecfreq}{\signalrxvec_{\B{f}}}

\newcommand{\mi}{\mathcal{I}}
\newcommand{\miuser}[1]{\mi_{\user}}

\newcommand{\speceff}{\mathcal{C}}
\newcommand{\specefful}{\speceff^{(\ul)}}
\newcommand{\speceffdl}{\speceff^{(\dl)}}
\newcommand{\speceffopt}{\speceff_{\opt}}
\newcommand{\speceffoptul}{\specefful_{\opt}}
\newcommand{\speceffoptdl}{\speceffdl_{\opt}}
\newcommand{\speceffsiso}{\speceff_{\text{siso}}}
\newcommand{\speceffuser}[1]{\speceff_{#1}}

\newcommand{\speceffkey}{\speceff_{\key}}
\newcommand{\speceffkeyul}{\specefful_{\key}}
\newcommand{\speceffkeydl}{\speceffdl_{\key}}

\newcommand{\capacity}{\speceff_{B}}
\newcommand{\capacityopt}{\capacity_{\opt}}

\newcommand{\speceffest}{\hat{\speceff}}
\newcommand{\speceffestnorm}{\hat{\speceff}_{\text{norm}}}
\newcommand{\speceffestfro}{\hat{\speceff}_{\text{fro}}}

\newcommand{\SNRmeas}[2]{\SNR{#1}{#2}^{(\text{meas})}}
\newcommand{\SNRref}[2]{\SNR{#1}{#2}^{(\text{ref})}}

\newcommand{\phaseangleopt}{\phi_{\opt}}
\newcommand{\phaseangleoptdiff}{\triangle\phaseangleopt}

\newcommand{\AtmMatrix}{\mymatrix{D}}

\newcommand{\userindex}{i}

\newcommand{\precodingmatrix}{\mymatrix{B}}
\newcommand{\precodingscaling}{\zeta}
\newcommand{\precodvec}[1]{\myvector{b}_{#1}}
\newcommand{\precodingmatrixul}{\precodingmatrix_{\ul}}
\newcommand{\precodingmatrixulordl}{\precodingmatrix_{\ul/\dl}}
\newcommand{\precodingmatrixdl}{\precodingmatrix_{\dl}}
\newcommand{\qmatrix}[1]{\mymatrix{Q}_{#1}}
\newcommand{\qmatrixblk}[1]{\mymatrix{\bar{Q}}_{#1}}

\newcommand{\eirpul}{P_{\ul}}
\newcommand{\eirpdl}{P_{\dl}}

\newcommand{\AntNumSLrefl}{\AntNumSL_{\text{refl}}}
\newcommand{\UserNum}{K}
\newcommand{\UserNumTot}{\UserNum_{\text{tot}}}
\newcommand{\numOfGroups}{N_\text{G}}

\newcommand{\user}{k}
\newcommand{\usergroupindex}{g}

\newcommand{\AntGainVec}[1]{\mymatrix{g}_{#1}}
\newcommand{\AntGain}[2]{g_{#1#2}}
\newcommand{\AntGainUL}[2]{g^{(\ul)}_{#1#2}}
\newcommand{\AntGainDL}[2]{g^{(\dl)}_{#1#2}}

\newcommand{\AntGainMax}{G_{\text{max}}}
\newcommand{\AntGainMatrix}{\mymatrix{G}}
\newcommand{\AntGainMatrixdl}{\mymatrix{G}_{\dl}}
\newcommand{\AntGainMatrixul}{\mymatrix{G}_{\ul}}

\newcommand{\CTV}[1]{\mymatrix{h}_{#1}}
\newcommand{\CTVdl}[1]{\mymatrix{h}_{\dl,#1}}
\newcommand{\CTVdltilde}[1]{\tilde{\mymatrix{h}}_{\dl,#1}}
\newcommand{\CTVul}[1]{\mymatrix{h}_{\ul,#1}}

\newcommand{\deltalon}{\triangle\lon}

\newcommand{\phaseRms}{\sigma_{\text{p}}}

\newcommand{\ScalingPrec}{\mu} 
\newcommand{\ScalingPrecVec}{\boldsymbol{\mu}} 
\newcommand{\ScalingPrecVecUL}{\boldsymbol{\mu}_{\ul}} 
\newcommand{\ScalingPrecVecDL}{\boldsymbol{\mu}_{\dl}} 
\newcommand{\ScalingPrecUL}[1]{\ScalingPrec_{\ul#1}} 
\newcommand{\ScalingPrecDL}[1]{\ScalingPrec_{\dl#1}} 


\author{Robert~T.~Schwarz,~\IEEEmembership{Member,~IEEE,} Thomas~Delamotte,~\IEEEmembership{Member,~IEEE,} Kai-Uwe~Storek,~\IEEEmembership{Member,~IEEE,} and~Andreas~Knopp,~\IEEEmembership{Senior Member,~IEEE}%
\thanks{Manuscript submitted May 30, 2018; revised August 8, 2018, and October 18, 2018.}
\thanks{R. T. Schwarz, T. Delamotte, K.-U. Storek, and A. Knopp are with the Chair of Signal Processing, Bundeswehr University Munich, 85579 Neubiberg, Germany (e-mail: papers.sp@unibw.de).}%
}

\markboth{IEEE Transactions on Broadcasting}%
{Accepted paper}

\maketitle

\begin{abstract}
\Acrlongpl{hts} employing multibeam antennas and full frequency reuse for broadband satellite services are considered in this paper. Such architectures offer, for example, a cost-effective solution to optimize data delivery and extend the coverage areas in future 5G networks. We propose the application of the \gls{mimo} technology in both the feeder link and the multiuser downlink. Spatial multiplexing of different data streams is performed in a common feeder beam. In the user links, \gls{mimo} with multiple beams is exploited to simultaneously serve different users in the same frequency channel. Under particular design constraints, effective spatial separation of the multiple user signals is possible. To mitigate the inter-stream interference in the \gls{mimo} feeder link as well as the multiuser downlink, precoding of the transmit signals is applied. Simulation results illustrate the performance gains in terms of sum throughput. 
\end{abstract}

\begin{IEEEkeywords}
MIMO, satellite communication, channel capacity, multiuser channels, radiowave propagation, 5G, precoding, user scheduling, non-terrestrial networks, backhauling, access networks
\end{IEEEkeywords}

\section{Introduction}
\label{sec:introduction}

\glsresetall

Broadcasting of video content to home users has been the key application scenario of geostationary satellite systems for many decades. \replaced[id=td]{Meanwhile, the advent of 5G networks and the introduction of integrated satellite-terrestrial architectures will considerably change the role of satellite communications in the near future. Traffic offloading to the network edges, backhauling or direct broadband access (e.g., \acrlong{vod}) to remote areas belong to the most promising use cases of \gls{satcom} \cite{Satcom5G}. Other use cases include the delivery of broadband data to satellite \glspl{esomp} like trains, cruise ships and airplanes. The cost-per-bit is in this context a key enabler.}{The advent of the Internet at anytime along with the introduction of \gls{vod} services has considerably changed the user's expectations how multimedia contents shall be delivered to the households. Individually selectable content at any time is the current state-of-the art, which requires the implementation of unicast transmission capabilities in future satellite systems in order to compete with terrestrial telecommunications systems. In addition, high-definition and, more recently, ultra-high-definition multimedia contents have dramatically increased the data rate demands to be supported by modern satellite systems for \glspl{fss}.} 
The sustaining demand for higher data rates in \replaced[id=td]{next-generation networks}{broadband \glspl{fss}} has already motivated the development of \glspl{hts}. \deleted[id=rs]{Key technologies of \gls{hts} systems are multibeam antennas supporting hundreds of geographically separated user beams and the reuse of the frequency spectrum.} 

\replaced[id=rs]{To}{In order to} meet the target data rates of future \gls{hts} systems \added[id=rs]{and the economies of scale in terms of costs per bit as required by 5G applications and networks, the trend has been to increase the number of beams to provide higher power flux density per beam and increase the reuse factors for the spectrum.} \deleted[id=rs]{, a \gls{ffr} strategy has recently been considered \mbox{\cite{Vazquez2016}}.} In contrast to \replaced[id=rs]{a}{the contemporary} \gls{fr4} scheme, where orthogonality between the adjacent beams is ensured using disjoint frequencies and polarizations, \added[id=rs]{the \gls{ffr} of the spectrum has recently been considered \cite{Vazquez2016}}. \added[id=rs]{However, }\gls{ffr} leads to significant inter-beam interference, also called \gls{cci},\added[id=rs]{ and moves the power-limited link budget to an interference-limited regime}. Users located at the edges of the beams suffer from the most severe \gls{cci}\replaced[id=rs]{ and experience}{. As a consequence, these users experience} a strong degradation of their achievable \gls{cinr}.

One strategy to significantly reduce the \gls{cci} is the precoding of the transmit signals in the gateways. In general, linear and nonlinear precoding techniques can be considered. While nonlinear precoders are hard to implement in practical systems, thus serving only as a theoretical upper limit in many studies, linear precoding techniques often achieve similar performances like their nonlinear counterparts \cite{Christopoulos2012}. Remarkable throughput improvements compared to the widespread \gls{fr4} scheme can be achieved with linear precoding strategies like for example \gls{zf} \cite{Arnau2012}. \deleted[id=td]{More recent}\added[id=td]{Recent} research activities in \cite{Christopoulos2015} and \cite{Joroughi2017} have a special focus on precoding for multicast communication\deleted[id=td]{in order} to exploit potential throughput gains offered by the DVB-S2X superframes \cite{EuropeanTelecommunicationsStandardsInstitute2014}. 

To support the huge amount of aggregated user traffic, the shift of the feeder links to the Q/V-band offering unexplored spectrum of up to \SI{5}{GHz} in bandwidth has recently been considered \cite{Cola2015}. Nevertheless, several gateways are still necessary \cite{Kyrgiazos2014b}. The gateways have to be displaced geographically to allow the reuse of the feeder link bandwidth. Moreover, transmit diversity techniques are needed to cope with the heavy rain fades in these frequency bands \cite{Siles2015}. Different solutions can be found in the literature, e.g. in \cite{Gharanjik2015} and the references therein. All solutions are based on a certain amount of redundancy in the feeder links, either through additional gateways in standby that become active if one gateway is in outage, or through spare capacity which is reserved at the active gateways. For traffic re-routing, a terrestrial network to interconnect the gateways is presumed in all cases, which is already common practice in current systems.

Relying on the existing gateway infrastructures as well as resorting to the \gls{ffr} strategy in the user beams, we propose the application of spatial \gls{mimo} techniques to tackle the uplink and downlink bandwidth limitations \replaced[id=rs]{and to reduce the per-bit delivery costs}{of \gls{hts} systems}. \Gls{mimo} systems are well-known for their \replaced[id=rs]{bandwidth efficiency}{spatial multiplexing gain} through the simultaneous transmission of multiple data streams in the same frequency band. \added[id=rs]{This is among the most important performance indicators to compete with terrestrial Gigabit networks.} Under favorable channel conditions, the spatial interference is eliminated in the channel, and a linear increase of the channel capacity with the number of transmit or receive antennas, whatever number is smaller, is achieved.

In \gls{mimo} channels with predominant \gls{los} wave propagation, \deleted[id=rs]{e.g. the \gls{fss} channel,} particular antenna array geometries are required \cite{Driessen1999} \deleted[id=td]{in order }to obtain such favorable channel conditions\deleted[id=td]{ to benefit from spatial multiplexing}. 
\deleted[id=rs]{In \mbox{\cite{Schwarz2008}}, }The optimal antenna array geometry for maximum-capacity \gls{los} \gls{mimo} satellite channels has been analytically derived\added[id=rs]{ in \cite{Schwarz2008}}\deleted[id=rs]{. 
After years of intensive research on the opportunities and limitations imposed by practical constraints, the theoretical approach from \mbox{\cite{Schwarz2008}}} and was\deleted[id=rs]{finally} verified through satellite channel measurements reported in \cite{Storek2016b,Hofmann2016,Hofmann2017}. Based on these fundamentals, very recent research has now started to propose first practical applications of \gls{mimo} to \gls{satcom} \cite{Storek2017}. \replaced[id=rs]{Here, the distribution of \gls{vod} services as an example of edge content delivery to home users in future 5G satellite networks is addressed\mbox{\cite{Giambene2018}}. However, the results can also be further extended to other \gls{fss} applications like the \glspl{esomp}.}{In that respect, this paper is dedicated to satellite broadcast.} 

In this paper we consider the forward link of a \gls{hts} system. In particular, we propose a \gls{mimo} feeder uplink with two gateway antennas and a \gls{mu-mimo} downlink to noncooperative users with a single antenna each. The satellite payload relies on a bent-pipe architecture \cite[Chapter 9.2, p. 437]{Maral2009} in order to keep the complexity at satellite level low. The content delivery to the users is achieved through spatial multiplexing of the different data streams.

\deleted[id=td]{Relying on the results of \mbox{\cite{Schwarz2008}}, an analytical expression for the optimal geographical positioning of the gateway antennas in an \gls{hts} scenario is derived in this paper. %
In the downlink a smart grouping of the geographically separated users is applied \mbox{\cite{Storek2017}} in order to form a favorable \gls{mimo} downlink channel that minimizes the \gls{cci} and maximizes the individual \glspl{cinr}. The remaining spatial interference in the \gls{mimo} uplink and the \gls{mu-mimo} downlink is mitigated with precoding at the gateways. This allows low complexity receivers at the \glspl{ut}.} 

\deleted[id=td]{Applying \gls{mimo} in the uplink and the downlink is completely different from existing beam interference mitigation strategies because the interference is actively exploited in order to enable a spatial multiplexing strategy. This approach can be seen as an amendment to the known \gls{cci} mitigation strategies that mostly work in frequency or time, trying to suppress inter-beam interference by precoding \mbox{\cite{Arnau2012}}. For the first time we address both, the application of spatial \gls{mimo} in the uplink in combination with \gls{ffr} in a multiuser \gls{mimo} downlink.}

\added[id=td]{The use of spatial \gls{mimo} in the uplink and in the downlink is a completely different approach compared to satellite network architectures that have been proposed so far. In state-of-the-art feeder links, the deployment of antennas separated by several tens of kilometers is only done to achieve diversity gains within the footprint of a given feeder beam. Both antennas are never active at the same time, which is not a cost-efficient solution due to the unused redundant hardware. With \gls{mimo} feeder links, the antennas are operated simultaneously, which doubles the maximum transmit power per link and enables spatial multiplexing. The opportunities of \gls{mimo} feeder links have been discussed by the authors in \cite{Delamotte2018b}.}

\added[id=td]{In the \gls{mu-mimo} downlink, the exploitation of the \gls{los} channel phase information allows to build groups of users with limited \gls{cci}. These users are spread over an area covered by different beams and are scheduled within the same time slots. The proposed \gls{mimo} solution introduces a novel philosophy that does not try to arbitrarily allocate users to a predefined beam according to their position in the multibeam coverage as suggested in known schemes \cite{Vazquez2016}. That way, in a \gls{mimo} \gls{ffr} scheme, the beam pattern is resolved and reduced to nothing more than a shaping of the power flux density on Earth. Such a beamfree approach was first proposed in \cite{Storek2017}. In a further step, this new concept might be used as another degree of freedom for throughput optimization, even leading to dynamically adaptable shaping depending on the capabilities of the satellite antenna multibeam architecture. In this work, an innovative \gls{hts} system design with a joint optimization of a smart gateway relying on spatial \gls{mimo} together with a novel scheduling algorithm in the user links is addressed.}

In order to follow the proposed ideas for the novel \gls{mimo} \gls{hts} application, some basic findings on \gls{mimo} for \replaced[id=rs]{\gls{los} satellite channels}{\gls{fss}} are needed. \deleted[id=rs]{Although available in the literature, these theoretical basics are inconveniently spread around different conference publications. Therefore, along with the novel \gls{hts} application, we will take this opportunity to also provide the first comprehensive presentation of the basic theory on \gls{mimo} for \gls{los} satellite channels in a thorough and rigorous manner. }
\deleted[id=rs]{We extend the result of our previous publication \mbox{\cite{Schwarz2008}} and derive a new analytical expression for the optimal positioning of the MIMO antenna elements in an \gls{hts} \gls{fss} scenario. This new expression allows to calculate the optimal antenna positions onboard a single MIMO satellite with multiple antennas as a uniform linear array. While the solution in \mbox{\cite{Schwarz2008}} was limited to two antennas in the orbit, we extend the design approach to arbitrary antenna numbers.} 
Section \ref{sec:mimosatcombasics} has been reserved for this aspect. 
The proposed \gls{mimo} \gls{hts} system is then thoroughly described in Section \ref{sec:mimohtsexample}, and the performance is assessed in terms of the sum throughput in Section \ref{subsec:SimulationResults}. Section \ref{sec:conclusion} concludes the paper.

\textbf{Notation:}\deleted[id=td]{Throughout the paper we use the following mathematical notations:} $\myeye{M}$ denotes the $M\times M$ identity matrix and $\myvector{1}$ is an all-ones vector of proper dimension. Operator $(.)^{\conj}$ denotes the complex conjugate while $(.)^{\Tr}$ and $(.)^{\He}$ denote the transpose and the complex conjugate transpose of a matrix or a vector. The functions $\mydiag{.}$ and $\mytrace{.}$ abbreviate the diagonal and trace operators. The notation $\mynorm{.}$ represents the Euclidean vector norm and $\myabs{.}$ gives the absolute value of a scalar. The operator $\myminvalue{.}$ (or $\mymaxvalue{.}$) returns the minimum (or maximum) value. The symbol $\odot$ is the Hadamard product, i.e. the element-wise multiplication of two matrices, and the operator $\left\lfloor x\right\rfloor$ gives the greatest integer less than or equal to $x$. Finally, $\left(.\right)^{\mpi}$ represents the Moore-Penrose pseudo-inverse of a matrix.

\section{\gls{mimo} \gls{satcom} Basics}
\label{sec:mimosatcombasics}

\renewcommand{\countNumSL}{n}
\renewcommand{\AntNumSLtx}{N}

In this section, the basics on \gls{mimo} over satellite for \gls{fss} are summarized. The focus is on the \gls{mimo} satellite channel and its correct modeling as a prerequisite for reliable system performance predictions. The channel capacity is used as a measure to assess the channel properties. We will show that the signal phase is a key property of the \gls{em} waves that needs to be considered in order to obtain high \gls{mimo} gains. We introduce the \gls{mimo} free space propagation model and neglect atmospheric effects for now. However, Section \ref{subsec:ChannelImpairments} will later be solely devoted to the discussion of atmospheric effects. Since the free space medium is isotropic, the channel is reciprocal and it is, therefore, sufficient to initially concentrate on the downlink. 
The presented results provide the necessary fundamentals for the discussion of the proposed \gls{mimo} \gls{hts} system architecture covering both the uplink and the downlink channel.

\subsection{Free Space \gls{mimo} \gls{satcom} Channel Model}
\label{subsec:channelmodel}

We consider a \gls{mimo} satellite downlink for \gls{fss} between $\AntNumSLtx$ satellite transmit antennas in the \gls{geo} and $\AntNumRx$ earth station receive antennas. 
The satellite acts as the \gls{mimo} transmitter while the fixed earth station is the \gls{mimo} receiver. The vector $\signalrxvec=\left[\signalrxvecentry{1},\ldots,\signalrxvecentry{\AntNumRx}\right]^{\Tr}$ is the vector of receive signals at a given time instance with $\signalrxvecentry{\countNumRx}$ denoting the signal in complex baseband notation at the $\countNumRx$-th receive antenna. $\signalrxvec$ is calculated as
\begin{equation}
	\signalrxvec=\CTM\signaltxvec{}+\noisevec{},
\label{eq:rxvector}
\end{equation}
with $\signaltxvec{}=\left[\signaltxvecentry{1},\ldots,\signaltxvecentry{\AntNumSLtx}\right]^{\Tr}$ and $\noisevec{}=\left[\noisevecentry{1},\ldots,\noisevecentry{\AntNumRx}\right]^{\Tr}$. The symbols $\signaltxvecentry{\countNumSL}$ and $\noisevecentry{\countNumRx}$ are the transmit symbol at the $\countNumSL$-th satellite antenna and the noise contribution at the $\countNumRx$-th receive antenna, respectively. 
The noise entries are \gls{iid} circularly symmetric complex Gaussian variables which are uncorrelated with the data symbols. The matrix $\CTM\in\mathds{C}^{\AntNumRx\times\AntNumSLtx}$ denotes the \gls{mimo} channel matrix. The calculation of its entries is detailed in the following.

We focus on frequency bands well above \SI{10}{GHz}. In those frequency bands, high-gain and directive antennas are required to obtain a sufficient link budget and close the link with high throughput. Moreover, narrow main beams with low side lobe levels effectively suppress interfering signals from and to neighboring satellite systems, and are, therefore, a design objective for earth station antennas operating with \gls{geo} satellites \cite{ITURS580}. Relying on such typical earth station antennas, it is assumed that any multipath contributions are suppressed by the directional antennas. Neglecting the atmospheric effects, the satellite channel can be described using a deterministic \gls{los} model based on the free space wave propagation.

The \gls{los} channel coefficient $\CTMentry{\countNumRx}{\countNumSL}$ between the $\countNumSL$-th satellite antenna and the $\countNumRx$-th earth station antenna, which corresponds to the $(\countNumRx,\countNumSL)$-th entry of $\CTM$, is given in the equivalent baseband notation by
\begin{equation}
	\CTMentry{\countNumRx}{\countNumSL} = \FSL{\countNumRx}{\countNumSL} \cdot \mye{-}{\frac{2\pi}{\carrierwavelength}\distanceant{\countNumRx}{\countNumSL}} \approx \FSL{}{} \cdot \mye{-}{\frac{2\pi}{\carrierwavelength}\distanceant{\countNumRx}{\countNumSL}}.
	 \label{eq:loschannelcoefficient}										
\end{equation} 
Here, $\carrierwavelength=\speedoflight/\carrierfreq$ denotes the wavelength of the carrier with frequency $\carrierfreq$, $\speedoflight$ is the speed of light, and $\FSL{\countNumRx}{\countNumSL} = \carrierwavelength/\left(4 \pi \distanceant{\countNumRx}{\countNumSL}\right) \cdot \e^{\imj\channelphase}$ models the free space propagation gain. The parameter $\channelphase$ stands for the common carrier phase and can be assumed to be zero \gls{wlog}. The parameter $\distanceant{\countNumRx}{\countNumSL}$ 
denotes the distance between the $\countNumSL$-th satellite transmit antenna and the $\countNumRx$-th earth station receive antenna. On the right hand side of (\ref{eq:loschannelcoefficient}), we applied the approximation $\FSL{\countNumRx}{\countNumSL}\approx\FSL{}{}=\carrierwavelength/\left(4 \pi \meandistanceant\right)$ with $\meandistanceant=1/\left(\AntNumRx\AntNumSLtx\right)\cdot\sum_{\countNumRx=1}^{\AntNumRx}\sum_{\countNumSL=1}^{\AntNumSLtx} \distanceant{\countNumRx}{\countNumSL}$. This is reasonable because the difference between the path lengths is very small compared to their mean total length.\footnote{To give an example: Assume that one earth station antenna is located at the sub-satellite point while a second earth station antenna has a relative distance to the first earth station antenna of \SI{3}{\degree} in geographical longitude (corresponds to a distance of approximately \SI{340}{km}). In this case the relative error is approximately \num{2.9e-4}. In other words, the magnitude of the amplitude has an error of \SI{0.04}{dB} at \SI{20}{GHz}, which can be neglected.} 
Note again that this model will be extended in subsection \ref{subsec:ChannelImpairments} to additionally take relevant atmospheric effects into account.

It is also important to note that ray tracing through the parameter $\distanceant{\countNumRx}{\countNumSL}$ has been applied to exactly determine the phase entries of $\CTM$. This is referred as the \gls{swm} in the literature and stands in contrast to the \gls{pwm} which assumes no relevant phase differences between the entries of $\CTM$ \cite{Jiang2005}. As shown in the following, the application of the \gls{swm} is a fundamental prerequisite to correctly forecast the capacity provided by a \gls{mimo} satellite system and to derive the relevant design criteria for its capacity optimization. \deleted[id=td]{This will be detailed in the next section.}\added[id=td]{This is detailed in the next section.} 

\subsection{\gls{mimo} Channel Capacity}
\label{subsec:capacity}

Consider again the downlink scenario, in which the single-satellite is the \gls{mimo} transmitter and the earth station is the \gls{mimo} receiver. 
Based on the deterministic \gls{los} model in (\ref{eq:loschannelcoefficient}), the time invariant \gls{mimo} channel capacity without channel knowledge at the transmitter is given by \cite{Telatar1999}
\begin{equation}
	\speceff=\mylog{2}{\mydet{\myeye{\AntNumRx}+\SNR{}{}\cdot\CTM\CTM^{\He}}}.
\label{eq:capacity}
\end{equation}
Here, $\SNR{}{}$ is the \gls{cnr}, which is defined as the ratio of the transmit power per satellite antenna to the noise power per earth station receive antenna. 

To illustrate the dependence of the channel capacity on the properties of $\CTM$, let us decompose the \gls{mimo} channel into parallel sub-channels, so called eigenmodes, with a \gls{svd} of the form $\CTM=\Umatrix\Gammamatrix\matrixV^{\He}$. Moreover, to ease the mathematical notation, $\maxNumAnt=\mymaxvalue{\AntNumRx,\AntNumSLtx}$ and $\minNumAnt=\myminvalue{\AntNumRx,\AntNumSLtx}$ are introduced. The matrices $\Umatrix$ and $\matrixV$ are both unitary and constitute an orthonormal basis of the column and row spaces of the channel matrix $\CTM$, respectively. The matrix $\Gammamatrix$ is a rectangular diagonal matrix with $\minNumAnt$ non-negative singular values $\sqrt{\eigenvalue{1}},\ldots,\sqrt{\eigenvalue{\minNumAnt}}$ of $\CTM$, sorted in descending order on the main diagonal. 
Using this decomposition, we obtain the equivalent model
\begin{equation}
    \tilde{\signalrxvec}=\Gammamatrix\tilde{\signaltxvec{}}+\tilde{\noisevec{}},
    \label{eq:parallelsiso}
\end{equation}
with $\tilde{\signalrxvec}=\Umatrix^{\He}\signalrxvec$, $\tilde{\signaltxvec{}}=\matrixV^{\He}\signaltxvec{}$ and $\tilde{\noisevec{}}=\Umatrix^{\He}\noisevec{}$. This way, the \gls{mimo} system from (\ref{eq:rxvector}) is transformed into $\minNumAnt$ parallel and non-interfering \gls{siso} channels. 

The channel capacity of this system is calculated by the sum over all parallel sub-channels, i.e.
\begin{equation}
    \speceff=\sum\limits_{\counteru=1}^{\minNumAnt}\mylog{2}{1+\SNR{}{}\eigenvalue{\counteru}},
    \label{eq:capacitySum}
\end{equation}
where $\eigenvalue{\counteru}$ are the eigenvalues of $\CTM\CTM^{\He}$. They equal the square of the singular values of $\CTM$. Since $\CTM\CTM^{\He}$ is positive semi-definite, the eigenvalues are in the range of %
$0\leq\eigenvalue{\counteru}\leq\mytrace{\CTM\CTM^{\He}}=\sum_{\counteru=1}^{\minNumAnt}\eigenvalue{\counteru}=\minNumAnt\maxNumAnt\myabs{\FSL{}{}}^2.$ 
The magnitude of the $\counteru$-th singular value in $\Gammamatrix$ represents the channel gain of the $\counteru$-th equivalent \gls{siso} channel or eigenmode. 

To explain the condition for which the maximum \gls{mimo} channel capacity is achieved, let us rewrite (\ref{eq:capacitySum}) as $\speceff=\sum_{\counteru=1}^{\minNumAnt}\mylog{2}{1+\SNR{}{}\eigenvalue{\counteru}}=\mylog{2}{\prod_{\counteru=1}^{\minNumAnt}\left( 1+\SNR{}{}\eigenvalue{\counteru}\right)}$. Since the logarithm is monotonically increasing, $\speceff$ is maximized by maximizing $\prod_{\counteru=1}^{\minNumAnt}\left( 1+\SNR{}{}\eigenvalue{\counteru}\right)$. It can be shown by basic algebra that, if the sum of $\minNumAnt$ non-negative numbers is fixed, their product is maximized for the case where they are all equal. Therefore the \gls{mimo} capacity is maximized when all eigenvalues are equal. Since $\sum_{\counteru=1}^{\minNumAnt}\eigenvalue{\counteru}=\minNumAnt\maxNumAnt\myabs{\FSL{}{}}^2$, the optimal eigenvalue profile must be $\eigenvalue{\counteru}=\maxNumAnt\myabs{\FSL{}{}}^2\,\forall\counteru$. In this case, a \gls{cnr} gain of $\maxNumAnt$ and a multiplexing gain of $\minNumAnt$ is achieved, and (\ref{eq:capacitySum}) yields 
\begin{equation}
	\speceffopt = \minNumAnt\mylog{2}{1+\SNR{}{}\maxNumAnt\myabs{\FSL{}{}}^2}.
\label{eq:capacitymax}
\end{equation}

From (\ref{eq:capacitymax}), it is clear that the multiplexing gain is limited to $\minNumAnt$, i.e. the maximum number of parallel sub-channels of $\CTM$. Therefore, an additional antenna at the link end with $\maxNumAnt$ antennas does not increase the multiplexing gain, but increases the \gls{cnr} by a factor of $\maxNumAnt+1$. However, as long as all $\minNumAnt$ eigenvalues are equal, the channel capacity is maximized according to (\ref{eq:capacitymax}), and $\CTM$ is called an ``optimal \gls{mimo} channel''. For $\AntNumRx>\AntNumSLtx$ (respectively, $\AntNumRx\leq\AntNumSLtx$), all column (row) vectors of $\CTM$ are then pairwise orthogonal and have equal norm. The latter is actually always fulfilled for the pure \gls{los} channel matrix since all the coefficients have identical magnitude. If $\AntNumRx=\AntNumSLtx$, $\CTM$ is a scaled unitary matrix having orthogonal row \emph{and} column vectors, i.e. $\CTM\CTM^{\He}=\CTM^{\He}\CTM=\AntNumRx\myabs{\FSL{}{}}^2\myeye{\AntNumRx}$. 

A simple $2\times 2$ example of an optimal \gls{mimo} channel is $\CTM=\FSL{}{}\left[ \begin{smallmatrix} 1&1\\ 1&\mye{-}{\pi} \end{smallmatrix} \right]$, which again highlights the need to model the different phase entries of $\CTM$. 
Here the two transmit signals impinge phase aligned at the first receive antenna, and they exhibit a phase difference of $\pi$ at the second receive antenna. A phase difference of $\pi$ corresponds to a difference among the path lengths of $\carrierwavelength/2$. The condition to obtain such channels will be derived in Section \ref{subsec:OptimalSatelliteChannel}, and it will be shown in Section \ref{subsec:AntennaPositions} that this condition requires particular spacings between the antennas at the transmitter and the receiver. 

Consider now the example where all receive signals at each receive antenna are nearly phase aligned, i.e. the \gls{pwm} can be applied. All path lengths are approximately equal and the channel matrix is $\CTM\approx\FSL{}{}\left[ \begin{smallmatrix} 1&1\\ 1&1 \end{smallmatrix} \right]$. This is the so-called ``keyhole channel'' \cite{Jiang2005}, which provides only one sub-channel or eigenmode. All eigenvalues are zero except of one that is $\eigenvalue{1}=\minNumAnt\maxNumAnt\myabs{\FSL{}{}}^2$. The keyhole capacity constitutes the lower capacity bound for \gls{mimo} systems and is given as
\begin{equation}
	\speceffkey = \mylog{2}{1+\SNR{}{}\minNumAnt\maxNumAnt\myabs{\FSL{}{}}^2}.
\label{eq:capacitykeyhole}
\end{equation}
\deleted[id=td]{It will be used for comparison purposes in the following.}\added[id=td]{It is used for comparison purposes in the following.}

\subsection{Optimal \gls{mimo} Satellite Channels}
\label{subsec:OptimalSatelliteChannel}

In this section, the general criterion for ``optimal'' \gls{mimo} satellite channels is derived. We have shown that the pure \gls{los} channel matrix $\CTM$ is optimal if all row (or column) vectors in case of $\AntNumRx\leq\AntNumSLtx$ (or $\AntNumRx>\AntNumSLtx$) are pairwise orthogonal. This requirement can be formulated as 
\begin{align}
\begin{aligned}
\CTV{\text{r},\counterk}\CTV{\text{r},\counterl}^{\He} &= 0,\,\counterk,\counterl\in\left\{1,\ldots,\AntNumRx\right\},~\counterk>\counterl\text{, if }\AntNumRx\leq\AntNumSLtx,\\
\CTV{\text{c},\counterk}^{\He}\CTV{\text{c},\counterl} &= 0,\,\counterk,\counterl\in\left\{1,\ldots,\AntNumSLtx\right\},~\counterk>\counterl\text{, if }\AntNumRx>\AntNumSLtx,\label{eq:ConditionOrthCTM}
\end{aligned}
\end{align}
where $\CTV{\text{r},\counterk}$ and $\CTV{\text{c},\counterk}$ denote the $\counterk$-th row vector and the $\counterk$-th column vector of $\CTM$, respectively. Both conditions in (\ref{eq:ConditionOrthCTM}) are equivalent, and it is, therefore, sufficient to consider the case $\AntNumRx\leq\AntNumSLtx$ in the following. We will see later that the result for $\AntNumRx>\AntNumSLtx$ is similar. 

Applying (\ref{eq:loschannelcoefficient}) to (\ref{eq:ConditionOrthCTM}) for $\AntNumRx\leq\AntNumSLtx$ we obtain\footnote{Please note that we divided (\ref{eq:ConditionOrthCTM}) by $\myabs{\FSL{}{}}^2$ to obtain (\ref{eq:ConditionOrthCTM2}).}
\begin{equation}
	\sum\limits_{\countNumSL=1}^{\AntNumSLtx} \mye{-}{\frac{2\pi}{\carrierwavelength}\left(\distanceant{\counterk}{\countNumSL}-\distanceant{\counterl}{\countNumSL}\right)} = 0,~\counterk,\counterl\in\left\{1,\ldots,\AntNumRx\right\},~\counterk>\counterl.
\label{eq:ConditionOrthCTM2}
\end{equation}
An appropriate choice of the various distances between the transmit and receive antennas is the key to satisfy (\ref{eq:ConditionOrthCTM2}) since the other parameters are constant for all $\AntNumSLtx$ phasors. 

To find a solution for (\ref{eq:ConditionOrthCTM2}), we first consider the following condition:
\begin{equation}
    \sum\limits_{\countNumSL=1}^{\AntNumSLtx}\mye{-}{\beta\left(\countNumSL+\kappa\right)}=0,~ \beta\in\mathds{R}\setminus\left\{0\right\},~\kappa\in\mathds{R}.
    \label{eq:CondRewritten}
\end{equation}
The sum in (\ref{eq:CondRewritten}) corresponds to the sum of $\AntNumSLtx$ terms of a geometric series and can be written as
\begin{equation}
\sum\limits_{\countNumSL=1}^{\AntNumSLtx}p\cdot q^{\countNumSL}=p\cdot q\cdot\frac{1-q^{\AntNumSLtx}}{1-q}=0,~p=\mye{-}{\beta\kappa},~q=\mye{-}{\beta}.
\label{eq:geometricseries}
\end{equation} 
The condition (\ref{eq:CondRewritten}) is satisfied if $1=\mye{-}{\beta\AntNumSLtx}$ and $1\neq\mye{-}{\beta}$. The solution that fulfills both constraints is 
\begin{equation}
	\beta=2\pi\vfactor{}/\AntNumSLtx,\,\vfactor{}\in\mathds{Z}\text{, with }\vfactor{}\nmid\AntNumSLtx,
\label{eq:solution1}
\end{equation}
where $\vfactor{}\nmid\AntNumSLtx$ means $\vfactor{}$ must not be a multiple of $\AntNumSLtx$. 
Using the substitution $\frac{2\pi}{\carrierwavelength}\left(\distanceant{\counterk}{\countNumSL}-\distanceant{\counterl}{\countNumSL}\right)=\beta\left(\countNumSL+\kappa\right)$ in (\ref{eq:CondRewritten}), we obtain 
\begin{align}
	\left(\distanceant{\counterk}{\countNumSL}-\distanceant{\counterl}{\countNumSL}\right) = \carrierwavelength\vfactor{\counterk\counterl}\left(\countNumSL+\kappa_{\counterk\counterl}\right)/\AntNumSLtx,\counterk,\counterl\in\left\{1,\ldots,\AntNumRx\right\},\label{eq:solution2}
\end{align}
with $\counterk>\counterl$, $\vfactor{\counterk\counterl}\in\mathds{Z}, \vfactor{\counterk\counterl}\nmid\AntNumSLtx$ and $\kappa_{\counterk\counterl}\in\mathds{R}$. 

Note that $\vfactor{\counterk\counterl}$ and $\kappa_{\counterk\counterl}$ can be different for different value pairs of $(\counterk,\counterl)$ because the $\AntNumSLtx$ phasors in (\ref{eq:ConditionOrthCTM2}) for one particular set of $(\counterk, \counterl)$ are independent of any other value pair $(\counterk', \counterl')$. The solution in (\ref{eq:solution2}) is the very general condition to obtain optimal \gls{mimo} satellite channels with arbitrary antenna number under \gls{los} conditions. Note that no particular constraints \gls{wrt} the geometrical arrangement of the antenna elements have been applied so far. 

If we assume the antennas are arranged as \glspl{ula}, (\ref{eq:solution2}) can be simplified to achieve the result reported in \cite{Sarris2006}. 
If \glspl{ula} at both link ends are applied and a large distance between the transmitter and receiver compared to the array dimensions is assumed,\footnote{This is a valid assumption for \gls{geo} applications since the Earth-to-space distance is at least \SI{35786.1}{km}, while the array dimensions are assumed to be not larger than several tens of km.} it can be revealed through geometrical analysis that 
\begin{equation}
	\left(\distanceant{\counterk}{\countNumSL+1}-\distanceant{\counterl}{\countNumSL+1}\right) - \left(\distanceant{\counterk}{\countNumSL}-\distanceant{\counterl}{\countNumSL}\right)\approx \const
\label{eq:MxN1}
\end{equation}
for $\countNumSL\in\left\{1,\ldots,\AntNumSLtx-1\right\}$ and $\counterk,\counterl\in\left\{1,\ldots,\AntNumRx\right\}$, $\counterk>\counterl$. Since (\ref{eq:MxN1}) holds for all combinations of the indices $\counterk,\counterl,\countNumSL$ with $\counterk>\counterl$, we can set \gls{wlog} $\counterk=2$, $\counterl=1$ and $\countNumSL=1$. Using (\ref{eq:solution2}) in (\ref{eq:MxN1}) finally yields
\begin{equation}
	\left(\distanceant{2}{2}-\distanceant{1}{2}\right) - \left(\distanceant{2}{1}-\distanceant{1}{1}\right) = \vfactor{}\carrierwavelength/\AntNumSLtx,\,\vfactor{}\in\mathds{Z},\,\vfactor{}\nmid\AntNumSLtx.
\label{eq:MxN2}
\end{equation}

This result has first been published in \cite{Schwarz2008} in 2008 for \gls{mimo} \gls{satcom} applications. Although the derivation has been slightly different and was limited to $\AntNumSLtx=2$ satellite antennas, (\ref{eq:MxN2}) tackles the solution presented in \cite{Schwarz2008}. It satisfies (\ref{eq:ConditionOrthCTM2}) under the constraint that \gls{ula} geometries are used at both link ends. Moreover, the result remains valid for $\AntNumRx>\AntNumSLtx$ by replacing $\AntNumSLtx$ with $\AntNumRx$ on the right hand side of the equation. If the differences between the path lengths satisfy (\ref{eq:MxN2}), the resulting \gls{mimo} \gls{los} channel exhibits the maximum channel capacity according to (\ref{eq:capacitymax}). Condition (\ref{eq:MxN2}) leads to particular requirements with respect to the positioning of the antenna elements as presented next. Moreover, the analysis shows that the signal phase must be taken into account.

\subsection{Optimal \gls{mimo} \gls{satcom} Antenna Positioning}
\label{subsec:AntennaPositions}

In the following, we apply the result of (\ref{eq:MxN2}) to obtain maximum-capacity \gls{mimo} \gls{satcom} links. To this end, we need to calculate the distances between the transmit-receive antenna pairs, which are determined by the geographical locations of the antenna elements on Earth and in space. A set of geometrical design parameters is introduced that exactly defines the geographical locations of the antenna elements. Moreover, the analysis shows that the signal phase must be taken into account. This is fundamentally different from the great bunch of existing publications on \gls{mimo} satellite systems, which apply a \gls{pwm} through narrow antenna spacing and can, therefore, never achieve a higher capacity than keyhole. 

\begin{figure}[t]
	\centering
	\footnotesize
	\includegraphics{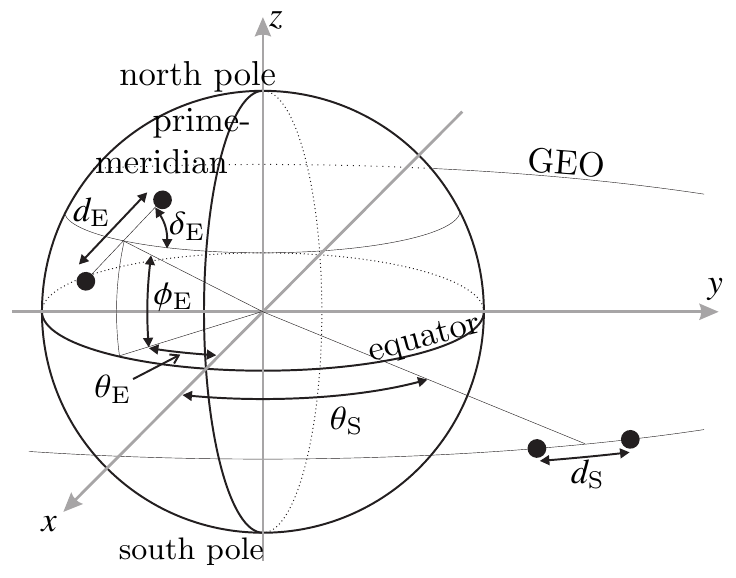}
	\caption{Parameter definition to describe the antenna positions, exemplary $\AntNumRx=2$ earth station antennas and $\AntNumSLtx=2$ satellite antennas are shown}
	\label{fig:parameterIntro}
\end{figure}

All the required parameters are illustrated in \mbox{Fig.\ \ref{fig:parameterIntro}}. The antenna locations are defined using an \acrfull{ecef} coordinate system. For the sake of simplicity, the distance between the Earth's center and any point on its surface is set equal to the mean Earth radius $\earthradius = \SI{6378.1}{\kilo\meter}$. Since the distance between the earth station antennas is small compared to $\earthradius$, the Earth curvature is neglected. It will be shown later in this section that this simplification has a negligible impact on the optimal design of the \gls{mimo} satellite link. The orientation $\ulaorientes$ characterizes the angle between the east-west direction and the antenna array. The pair of latitude $\lates$ and longitude $\lones$ specifies the center of the antenna array, and $\antspacinges$ is the inter-antenna distance. This allows to fully characterize the position of the ground antennas. The positioning vector of the $\countNumRx$-th earth station antenna in three-dimensional Cartesian coordinates is given in (\ref{eq:posesant}) at the top of the next page, where $\antspacingesant{\countNumRx}=\antspacinges\cdot\left(\countNumRx-1/2-\AntNumRx/2\right)$. 
\begin{figure*}[!t]
\normalsize
\setcounter{MYtempeqncnt}{\value{equation}}
\setcounter{equation}{15}

\begin{equation}
\posvecesant{\countNumRx}=
\begin{bmatrix}
 \earthradius\cos\lates\cos\lones - \antspacingesant{\countNumRx} \cdot \left( \sin\lones\cos\ulaorientes+\sin\lates\cos\lones\sin\ulaorientes \right)\\
 \earthradius\cos\lates\sin\lones + \antspacingesant{\countNumRx} \cdot \left( \cos\lones\cos\ulaorientes-\sin\lates\sin\lones\sin\ulaorientes \right)\\
 \earthradius\sin\lates + \antspacingesant{\countNumRx} \cdot \cos\lates\sin\ulaorientes
\end{bmatrix},~\countNumRx\in\left\{1,\ldots,\AntNumRx\right\},
\label{eq:posesant}
\end{equation}
\hrulefill
\end{figure*}

At the satellite, the antennas are considered to be positioned in the equatorial plane. Denoting $\lonsl$ as the longitude of the center of the antenna array, $\antspacingsl$ as the inter-antenna spacing, and $\orbitradius = \SI{42164.2}{\kilo\meter}$ as the ideal \gls{geo} radius, the position of the $\countNumSL$-th satellite antenna in three-dimensional Cartesian coordinates is given by 
\begin{multline}
    \posvecslant{\countNumSL}=\left[\orbitradius\cos\lonsl - \antspacingslant{\countNumSL}\sin\lonsl, \orbitradius\sin\lonsl\, + \right. \\ 
		\left. \antspacingslant{\countNumSL}\cos\lonsl, 0 \right]^{\Tr},
\end{multline}
with $\antspacingslant{\countNumSL} = \antspacingsl\cdot\left(\countNumSL-1/2-\AntNumSLtx/2\right)$. Here, an ideal \gls{geo} is assumed, i.e. the eccentricity and inclination of the satellite are negligibly small. The validity of this simplification will be justified in the remainder of this section. 

Based on the previous parametric characterization, the distance $\distanceant{\countNumRx}{\countNumSL}$ 
between the $\countNumRx$-th receive antenna and the $\countNumSL$-th satellite transmit antenna is then given by
\begin{align}
\distanceant{\countNumRx}{\countNumSL} = &\mynorm{\posvecesant{\countNumRx}-\posvecslant{\countNumSL}} = \distanceant{}{}\cdot\left( 1 + \subdelta{\countNumRx}{\countNumSL} \right)^{1/2}\text{, with }\label{eq:distance}
\end{align} 
\begin{multline}
	\subdelta{\countNumRx}{\countNumSL} = 2\left(\antspacingesant{\countNumRx} \orbitradius \anglesuba{} - 
	\antspacingslant{\countNumSL} \earthradius \anglesubb{} + 
	\antspacingesant{\countNumRx}\antspacingslant{\countNumSL} \anglesubc{}\right)/\distanceant{}{}^2 + \\
	\left(\antspacingesant{\countNumRx}^2 + \antspacingslant{\countNumSL}^2\right)/\distanceant{}{}^2.
	\label{eq:DELTA}
\end{multline} 
Here, $
\distanceant{}{} = \deltadistance\cdot\distancemin$ is the distance between the center of the earth station \gls{ula} and the satellite. Moreover, $\distancemin=\SI{35786.1}{km}$ is the minimum satellite-to-Earth distance, which is obtained if the earth station is directly located at the sub-satellite point. The parameter $\deltadistance=\left(1.42-0.42\cos\lates\cos\deltalon\right)^{1/2}$ with $1\leq\deltadistance\leq1.16$ describes the relative increase of the satellite-to-earth station distance depending on the geographical latitude $\lates$ and the relative longitude $\deltalon=\lones-\lonsl$. Furthermore, the substitutions $\anglesuba{} = \cos\ulaorientes\sin\deltalon + \sin\lates\sin\ulaorientes\cos\deltalon$, $\anglesubb{} = \cos\lates\sin\deltalon$, and $\anglesubc{} = \sin\lates\sin\ulaorientes\sin\deltalon - \cos\ulaorientes\cos\deltalon$ have been defined. 

Approximating the square root in (\ref{eq:distance}) by its second degree Taylor polynomial derived around $\subdelta{\countNumRx}{\countNumSL}=0$ provides 
\begin{equation}
	\distanceant{\countNumRx}{\countNumSL}\approx\, \distanceant{}{}\cdot\left(1+1/2\cdot\subdelta{\countNumRx}{\countNumSL} - 1/8\cdot\subdelta{\countNumRx}{\countNumSL}^2\right). 
\label{eq:approximation}
\end{equation}
Using (\ref{eq:approximation}) with (\ref{eq:DELTA}) in (\ref{eq:MxN2}) results in 
\begin{equation}
\antspacingsl\antspacinges / \distanceant{}{}\cdot\left(\anglesubc{}+0.21\anglesuba{}\anglesubb{}/\deltadistance^2\right) \approx \vfactor{}\cdot \carrierwavelength/\AntNumSLtx,\,\vfactor{} \in \mathds{Z},\,\vfactor{}\nmid \AntNumSLtx.\label{eq:FinalOptCondition}
\end{equation}
With respect to the desired accuracy of the Taylor approximation in (\ref{eq:approximation}), we require the absolute value of the total error $\myabs{\DeltaRemainder}$ of the left hand side of (\ref{eq:FinalOptCondition}) to be much smaller than the carrier wavelength $\carrierwavelength$. In particular, the residual shall be a fraction of the carrier wavelength only, for example $\myabs{\DeltaRemainder}\leq 1/100\cdot\carrierwavelength$, in order to obtain reliable results. Computer simulations have shown that in all practically relevant cases $\myabs{\DeltaRemainder}\leq\SI{3.8e-6}{m}$, which is sufficiently small to support carrier frequencies of more than \SI{100}{GHz}.

As expected, the two key parameters of the solution in (\ref{eq:FinalOptCondition}) are the antenna spacing $\antspacinges$ on Earth and $\antspacingsl$ in orbit. The spacing required is linearly proportional to the transmitter-receiver distance $\distanceant{}{}$ and the wavelength, i.e. $\antspacingsl\antspacinges\propto\distanceant{}{}\carrierwavelength$. Since $\distanceant{}{}\geq\SI{35786.1}{km}$, comparably large antenna spacings $\antspacingsl,\antspacinges$ are required to satisfy (\ref{eq:FinalOptCondition}). 

The minimum array dimensions are generally obtained if both antenna arrays are in broadside and the earth station is at the sub-satellite point. In this case, we have $\anglesubc{}=-1$, $\anglesuba{}=\anglesubb{}=0$, and $\distanceant{}{}=\distancemin$. The minimum spacing of the earth station array becomes $\antspacinges = \distancemin\carrierwavelength/\left(\AntNumSLtx\antspacingsl\right)$ if we set $\vfactor{}=1$. Smaller values of $\antspacinges$ lead to severe spatial interference because the receive array is no longer capable to spatially resolve each transmit antenna. All transmit signals can no longer be distinguished at the receiver and the \gls{mimo} channel converges to the \emph{keyhole channel} with $\CTM\approx\FSL{}{}\left[ \begin{smallmatrix} 1&1\\ 1&1 \end{smallmatrix} \right]$. A similar limit is also known as the Rayleigh criterion describing the resolution limit of optical systems. 

\begin{figure}[t]
	\centering
	\includegraphics{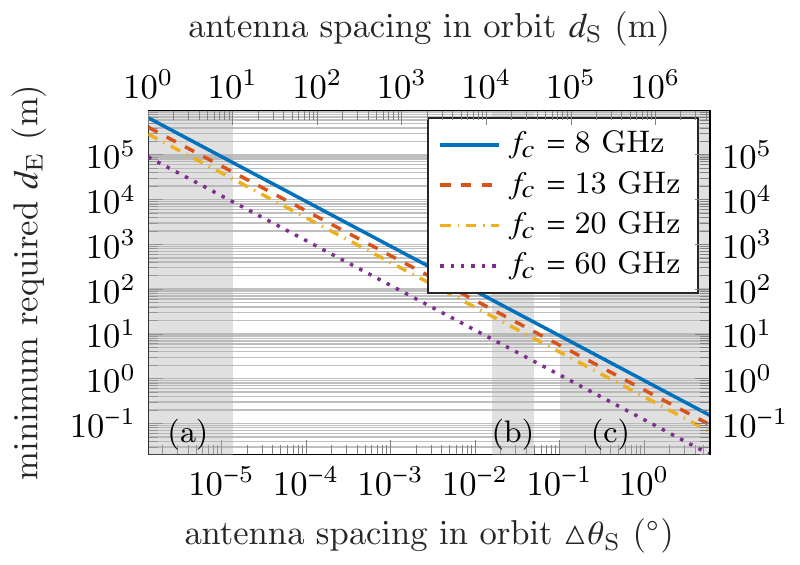}
	\caption{Minimum spacing $\antspacinges$ versus antenna spacing in orbit, shaded areas indicate useful ranges related to the categories: (a) single-satellite applications, (b) collocated satellite applications, (c) multiple-satellite applications}
	\label{fig:minantennadist}
\end{figure}

Fig. \ref{fig:minantennadist} shows the relation between $\antspacinges$ and the antenna spacing in orbit for different carrier frequencies. Since a wide range of values is provided the curves are shown in double-logarithmic scale. For large antenna separations in orbit, it is convenient to define the orbital separation $\spacingorbit$, given in degrees on the lower x-axis, while the upper x-axis shows $\antspacingsl$ in meters. The values are valid for $\AntNumRx=\AntNumSLtx=2$, but can easily be scaled to higher antenna numbers using the relation $\antspacinges\propto1/\AntNumSLtx$. Since $\anglesubc{}=-1$ and $\anglesuba{}=\anglesubb{}=0$, the antenna arrays are in broadside orientation, i.e. $\ulaorientes=\SI{0}{\degree}$, and $\deltalon=\SI{0}{\degree}$. As indicated by the shaded areas in Fig. \ref{fig:minantennadist}, we propose to classify \gls{mimo} \gls{satcom} systems into three basic categories. For each category a particular range of antenna spacings in orbit is basically feasible as follows:\newline
\textbf{(a) Single-Satellite Applications:} All \gls{mimo} antenna elements are on a single-satellite and the useful antenna spacing is in the range of $\SI{1}{m}\leq\antspacingsl\leq\SI{10}{m}$. A very promising and completely novel system proposal of this category is presented in the remainder of this paper.\newline
\textbf{(b) Collocated Satellite Applications:} Multiple satellites occupy a single orbital slot. Each satellite has one \gls{mimo} antenna element. A sufficient minimum separation between the spacecrafts must be ensured to account for inaccuracies of the tracking system and the thrusters \cite{Soop1994}. The upper bound is the station keeping window, which is typically $\pm\SI{0.05}{\degree}$ in longitude. We assume practically feasible antenna spacings to be in the range of $\SI{0.014}{\degree}\leq\spacingorbit\leq\SI{0.05}{\degree}$ (or equivalently $\SI{10}{km}\leq\antspacingsl\leq\SI{40}{km}$). Applications of this category are very similar to the single-satellite case but at increased complexity since novel spacecraft co-location strategies are required. 
\newline
\textbf{(c) Multiple-Satellite Applications:} Multiple satellites with one \gls{mimo} antenna each are located at different orbit positions resulting in a spacing of $\spacingorbit\geq\SI{0.1}{\degree}$. This category of applications requires non-directional antennas at the ground segment because directional antennas cannot point at different orbital slots at the same time. As a promising application, UHF \gls{satcom} has been proposed in \cite{Ramamurthy2016}\added[id=rs]{, and UHF \gls{mimo} satellite channel measurements reported in \cite{Hofmann2017} have shown a significant increase of the channel capacity}. \deleted[id=rs]{Moreover, measurements of the UHF \gls{mimo} satellite channel reported in \mbox{\cite{Hofmann2017}} have shown that the channel capacity can be significantly increased.} 
\newline

Fig. \ref{fig:minantennadist} emphasizes which spacing between the earth station antennas is at least required depending on the \gls{mimo} \gls{satcom} category considered. \replaced[id=rs]{I}{For example, i}n the single-satellite case, the minimum antenna spacing on Earth is approximately between \SI{10}{km} and \SI{100}{km}. Smaller antenna spacings $\antspacinges$ require larger spacings in the orbit, leading to collocated satellite applications or to multiple-satellite applications. Note that larger but still optimal values for $\antspacinges$ can be obtained if $\vfactor{}>1, \vfactor{}\nmid\AntNumSLtx$, is chosen because the optimal antenna spacing scales with $\vfactor{}\carrierwavelength/\AntNumSLtx,\,\vfactor{}\nmid\AntNumSLtx$.

The term $\left(\anglesubc{}+0.21\anglesuba{}\anglesubb{}/\deltadistance^2\right)\in\left[-1,+1\right]$ in (\ref{eq:FinalOptCondition}) can be interpreted as a \emph{reduction factor} because it apparently reduces the actually needed antenna spacings $\antspacinges$ and $\antspacingsl$, depending on the parameters $\ulaorientes$, $\deltalon$ and $\lates$. If the earth station \gls{ula} and the satellite \gls{ula} are not in broadside, we get $\myabs{\anglesubc{}+0.21\anglesuba{}\anglesubb{}/\deltadistance^2}<1$, and the antenna spacing has to be increased accordingly to still satisfy (\ref{eq:FinalOptCondition}). 
\begin{figure}[t]
  \centering
	\small
	\includegraphics{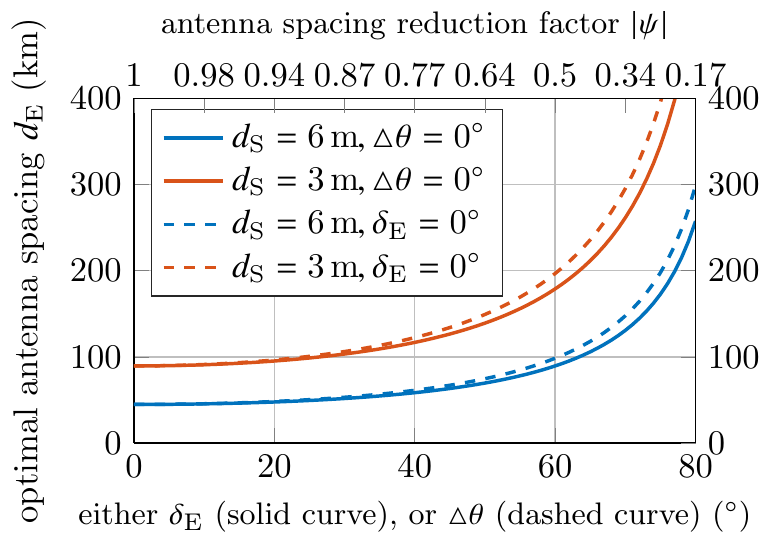}
  \caption{Optimal antenna spacing $\antspacinges$ \deleted[id=rs]{according to (\ref{eq:FinalOptCondition})} as a function of \gls{ula} orientation angle $\ulaorientes$ or relative longitude $\deltalon$ and $\anglesubc{}$, $\AntNumRx=\AntNumSLtx=2$, $\lates=\SI{0}{\degree}$, $\carrierfreq=\SI{20}{GHz}$ (single-satellite applications)}
	\label{fig:DegreeFreedomDiscuss}
\end{figure}

This required adjustment is shown in Fig. \ref{fig:DegreeFreedomDiscuss} with respect to $\antspacinges$. Two $2\times 2$ cases are considered: First, the earth station \gls{ula} is rotated by $\ulaorientes$ while the angles $\lates=\deltalon=\SI{0}{\degree}$ are fixed. This means that the earth station \gls{ula} is located at the sub-satellite point. This case corresponds to the solid curves, and the lower x-axis shows $\ulaorientes$ in degrees. The second case corresponds to the dashed curves and the relative longitude $\deltalon$ is increased while $\lates=\ulaorientes=\SI{0}{\degree}$ are fixed. In both cases $\anglesuba{}=\anglesubb{}=0$ and, thus, $\myabs{\anglesubc{}+0.21\anglesuba{}\anglesubb{}/\deltadistance^2}=\myabs{\anglesubc{}}$ as shown on the upper x-axis. In all cases, comparably large angular values of $\ulaorientes$ and $\deltalon$ are allowed for which the optimal spacing $\antspacinges$ remains approximately constant. This is due to the fact that the increment of $\antspacinges$ relates to the cosine of the respective angles. 

Taking the solid blue curve for $\antspacingsl=\SI{6}{m}$ as an example, the optimal $\antspacinges$ has to be increased by only about \SI{7}{km} compared to the minimum value of \SI{50}{km} for $\ulaorientes=\SI{30}{\degree}$, since $1/\myabs{\anglesubc{}}\cdot\antspacinges=1/0.87\cdot\SI{50}{km}=\SI{57}{km}$. The rotation of $\ulaorientes=\SI{30}{\degree}$ results in a \deleted[id=rs]{remarkable }displacement of the earth station antennas by \SI{12.5}{km} to the North and to the South. 

The increment of the dashed curves is slightly higher because the transmitter-receiver distance $\distanceant{}{}$ also increases with increasing $\deltalon$. However, assuming a fixed optimal earth station spacing of $\antspacinges=\SI{50}{km}$ at $\deltalon=\SI{0}{\degree}$ (blue dashed curve in Fig. \ref{fig:DegreeFreedomDiscuss}) and accepting \SI{10}{km} deviation from this optimum spacing to account for practical implementation constraints, a very large part of the \gls{geo} arc of more than $\deltalon=\pm\SI{30}{\degree}$ can be used. In other words, once an optimal setup has been implemented in terms of a ground station installation, it can be used for a wide range of satellite positions without significant
capacity degradation. \replaced[id=rs]{Therefore, }{This fact is important to note because it reveals that }the presented design constraint does not at all impose a flexibility disadvantage to the system.

\subsection{Sensitivity Discussion}

\begin{figure}
  \centering
	\includegraphics{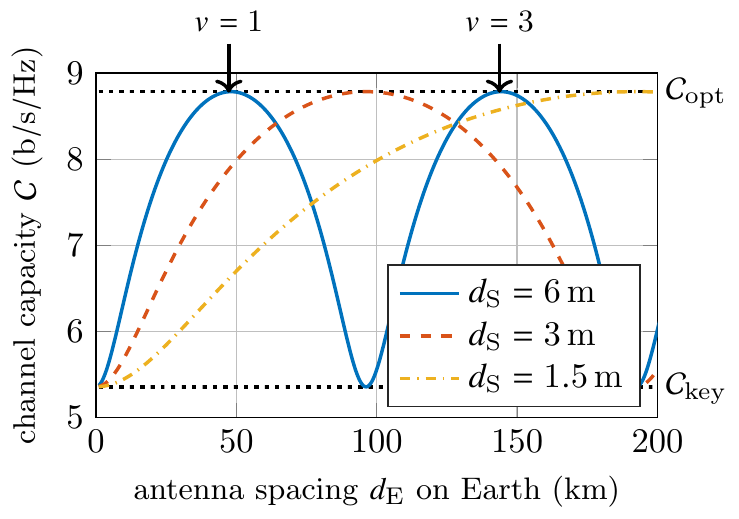}
  \caption{Capacity vs. antenna spacing on Earth for various antenna spacing at the satellite (single-satellite applications)}
	\label{fig:CapOverDist_SingleSat}
\end{figure}

To estimate the degradation of the \gls{mimo} capacity if the antennas are not optimally spaced, simulation results of $\speceff$ as a function of $\antspacinges$ are shown in Fig. \ref{fig:CapOverDist_SingleSat}. The curves relate again to the $2\times 2$ downlink at \SI{20}{GHz}. A receive \gls{cnr} of $10\log_{10}\SNR{}{}\myabs{\FSL{}{}}^2=\SI{10}{dB}$ is assumed. First, all curves show the expected periodic behavior. Each capacity maximum corresponds to one particular value of $\vfactor{}$. 

For example, the first and second maximum of the blue solid curve at approximately \SI{50}{km} and \SI{150}{km} correspond to $\vfactor{}=1$ and $\vfactor{}=3$, respectively. The keyhole capacity is obtained for $\vfactor{}=2$ at \SI{100}{km} because in this case $\vfactor{}$ is a multiple of $\AntNumSLtx=2$. Moreover, a large deviation of the optimal spacing $\antspacinges$ on Earth is indeed possible without a significant loss of $\speceffopt$. Taking again the blue solid curve for $\antspacingsl=\SI{6}{m}$ as an example, $\pm\SI{10}{km}$ around the first optimal value of \SI{50}{km} is allowed to still obtain \SI{8.6}{b/s/Hz} (the maximum is \SI{8.8}{b/s/Hz}). This is approximately \SI{98}{\%} of $\speceffopt$. The curves reveal that large deviations in all directions even in the kilometer-range can be accepted and still close-to-maximum capacity values are obtained. 

The same is true regarding possible displacements of the satellite in the \gls{geo}. Since orbit perturbations cause apparent displacements of the satellite \gls{wrt} its ideal geostationary position, the inclination and eccentricity vary over time. The amplitude of the overall orbital motion is specified by the station-keeping box, whose limits are usually $\pm\SI{0.05}{\degree}$ in longitude and latitude and \SI{4e-4}{} in eccentricity \cite{Maral2009}. 
In the single-satellite case, movements by this order of magnitude can generally be neglected. 
A displacement of the satellite in geographical longitude by \SI{0.05}{\degree} from its optimal position results in a variation of the relative longitude, i.e. $\deltalon\pm\SI{0.05}{\degree}$. 

The effect upon the optimal spacing of the earth station antennas has already been discussed by means of the dashed curves in Fig. \ref{fig:DegreeFreedomDiscuss}. The slope of the curves increases for larger values of the relative longitude $\deltalon$. Taking $\deltalon=\SI{80}{\degree}$ as an example, an adjustment of the earth station antenna spacing of \SI{8.2e-4}{m} would be required in order to compensate the satellite's longitude drift. Such values are far too small to have a remarkable impact on the channel capacity, so that the related effects can be neglected.  

In short, we conclude that for single-satellite applications:
\begin{itemize}
    \item The required positioning accuracy of the earth station antennas is manageable in practice. Several kilometers of antenna displacements in each direction are possible while still achieving very high, nearly optimal capacities.
    \item Satellite movements in the station-keeping window can be neglected.
    \item A large part of more than $\pm\SI{30}{\degree}$ of the \gls{geo} arc can be covered with the same earth station installation while limiting the capacity degradation to less than \SI{2}{\percent}.
\end{itemize}

The measurements reported in \cite{Hofmann2016} ultimately confirm the presented theory.\footnote{Since the measurement setups have been fairly complex, we do not repeat the results here, but we encourage the reader to take note of this practical field trial.} To conclude this introductory section about the basics of spatial \gls{mimo} over \gls{los} satellite channels, atmospheric impairments and the issue of differential signal delays need to be discussed. 

\subsection{Atmospheric Impairments and Further Aspects}
\label{subsec:ChannelImpairments}

\subsubsection{Atmospheric Impairments}
\label{subsubsec:AtmosphericImpairments}

In the frequency bands above \SI{10}{GHz}, the main radiowave propagation impairments originate from the troposphere and include attenuation effects as well as phase disturbances \cite{Allnutt2011}. An amplitude attenuation decreases the signal power, resulting in a lower \gls{cnr} for the affected antenna. Such a loss entails a capacity degradation similar to what would be observed in a \gls{siso} system suffering from the same attenuation\cite{Schwarz2009}. On the other hand, the impairments of the signal phase might affect the optimal phase relations within the channel matrix $\CTM$ and disturb the optimal eigenvalue profile. Fortunately, the channel capacity is not changed if the signal paths originating from the same antenna experience identical phase impairments \cite{Schwarz2009}. This latter property can be reasonably considered to be verified in practice. 

The assumption of identical phase impairments is based on the geometrical analysis of an optimal \gls{mimo} satellite link where the horizontal separation in the troposphere of two \gls{los} paths $\distanceant{\countNumRx}{1}$ and $\distanceant{\countNumRx}{2}$ is very small. In fact, the example in \cite[eq. (7)]{Schwarz2009} reveals a horizontal separation of less than \SI{1}{cm}. 
This theoretical assumption has been verified through interferometric measurements in the Ku-Band at \SI{12.5}{GHz} reported in \cite{Storek2016b}. The results prove that differential phase disturbances between neighboring \gls{los} paths can be neglected. 
Moreover, the long-term measurements have also shown that, once an optimal \gls{mimo} satellite link has been established, the maximum \gls{mimo} capacity can be obtained sustainably. It is, therefore, reasonable to model a common attenuation and phase shift for signal paths stemming from or arriving at the same earth station antenna. 

The atmospheric impairments for the $\countNumRx$-th earth station antenna are expressed as 
	$\AtmImp{\countNumRx} = \myabs{\AtmImp{\countNumRx}} \cdot \e^{-\imj\AtmPhase{\countNumRx}}$, 
where $\myabs{\AtmImp{\countNumRx}} \in \left[0,1\right]$ and $\AtmPhase{\countNumRx} \in \left[-\pi,\pi\right[$ represent respectively the additional amplitude attenuation and the phase shift. The attenuation in \si{\deci\bel} is obtained as 
\begin{equation}
	\AtmAtt{\countNumRx}=-20\cdot \mylog{10}{\myabs{\AtmImp{\countNumRx}}}.
\label{eq:attenuation}
\end{equation}
The \gls{los} free space channel coefficient in (\ref{eq:loschannelcoefficient}) is, thus, extended to %
$\CTMentrytilde{\countNumRx}{\countNumSL} = \CTMentry{\countNumRx}{\countNumSL}\cdot\AtmImp{\countNumRx}.$ 
It follows for the impaired channel transfer matrix 
\begin{equation}
    \CTMtilde = \Dmatrix\cdot\CTM,
    \label{eq:ImpairedCTM}
\end{equation}
with $\Dmatrix=\mydiag{\AtmImp{1},\ldots,\AtmImp{\AntNumRx}}$. The influence of the weather impairments on the \gls{mimo} feeder link of an \gls{hts} system will be analyzed in Section \ref{sec:mimohtsexample}.

\subsubsection{Differential Signal Delays}

A known issue of \gls{mimo} satellite links is the large difference in the propagation delay between the \gls{los} paths \cite{Horvath2006}. The time of arrival of the \gls{mimo} signals at the receiving antennas can vary by \deleted[id=rs]{tens or}hundreds of a symbol duration because of the large antenna spacing. This results in an asynchronous reception of those symbols which form a part of a single code word. 

To tackle this issue, a \gls{sc-fde} concept has been applied in \cite{Schwarz2011}. By using a sufficiently long guard interval, the different arrival times of the symbols can be compensated. 
The unavoidable loss in bandwidth efficiency depends on the length of the guard interval in relation to the frame length and usually does not exceed \SI{5}{\%} \cite{Schwarz2011}. An advantage of \gls{sc-fde} compared to other waveforms, which also rely on the use of a guard interval like \acrlong{ofdm}, is its very low \acrlong{papr}. 
However, modern waveforms currently discussed as 5G candidates \cite{Zhang2016a} are also potentially suitable to address these differential propagation delays in future systems.

\renewcommand{\countNumSL}{z}
\renewcommand{\AntNumSLtx}{Z_t}

\setlength\figureheight{7.5cm}
\setlength\figurewidth{15cm}
\section{\gls{mimo} \gls{hts} System Proposal}
\label{sec:mimohtsexample}

Relying on the result of Section \ref{sec:mimosatcombasics}, we now apply the \gls{mimo} concept to an \gls{hts} system. With an \gls{hts} system, the limitations of a broadcast scenario can be tackled and multiple users can be served with individual data streams. \Gls{hts} systems represent the most recent and powerful satellite architecture for high-data rate unicast and multicast communications. We will show how greatly these systems can benefit from spatial multiplexing in both their uplink and their downlink, to maximize the sum throughput. The considered \gls{fss} scenario belongs to the class of single-satellite \gls{mimo} applications since one \gls{hts} equipped with several antennas will be assumed.\footnote{The architecture of spatial \gls{mimo} in the feeder link of an \gls{hts} scenario has also been chosen in \cite{Delamotte2018a} and \cite{Delamotte2018b} for feeder link performance analyses that are out of scope of this paper.}

We recall that the objective of the considered illustrative example is to show the tremendous performance gain of a \gls{mimo} \gls{hts} system in terms of data rate if the \gls{mimo} \gls{los} system design approach from the previous section is considered. %
To this end, the proposed \gls{hts} system architecture with a \gls{mimo} feeder link and a \gls{mu-mimo} downlink is presented first together with the equivalent baseband model. The system model will be used in Section \ref{subsec:PrecoderDesign} to design the precoding strategies aimed at improving the spectral efficiency of the system. Section \ref{subsec:UserScheduling} describes the scheduling approach based on the algorithm in \cite{Storek2017} that allocates the resources to individual households. Simulation results are provided in Section \ref{subsec:SimulationResults} to finally assess the system performance in terms of sum rates in comparison to the state-of-the-art.

\subsection{System Description}
\label{subsec:SystemDescription}

\begin{figure*}[t]
	\centering
	\subfloat[]{{	\includegraphics{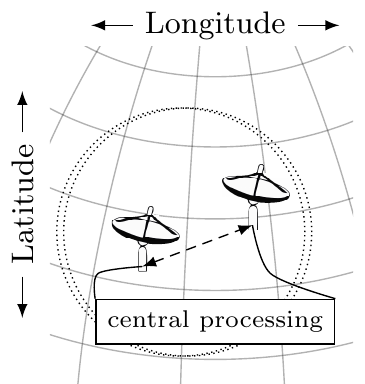}
	}
	\label{fig_first_case}}
	\hfil
	\subfloat[]{{\includegraphics{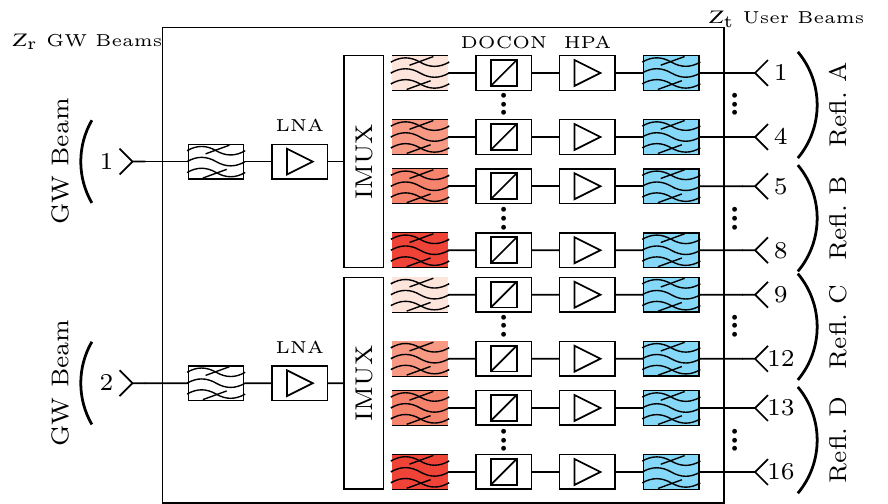}
	}
	\label{fig_second_case}}
	\hfil
	\subfloat[]{{\includegraphics{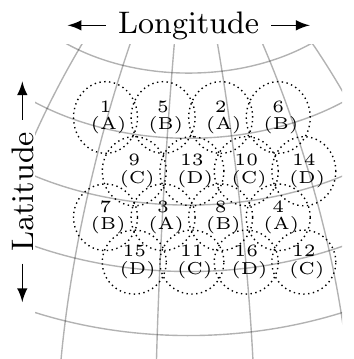}
	}
	\label{fig_third_case}}
	\caption{Proposed system architecture: (a) Two spatially separated gateway antennas are fed with signals by a common gateway station (central processing). (b) Schematic diagram of the proposed payload: Both gateway (GW) beams share a common frequency resource (V-band). The payload is designed to translate all incoming frequency bands to the same frequency band in the Ka-band in order to enable \gls{ffr}. The colors of the bandpass filters indicate the corresponding frequency range according to Fig.\ \ref{fig:freqPlanMimosiso}. (c) Geographical illustration of the 16 spot beam footprints. The dotted lines represent the \SI{3}{dB} contours of the multiple beams. The source feed and reflector are labeled inside every footprint.}
	\label{fig:HTS_SystArch}
\end{figure*}

\begin{figure}
	\centering
	\includegraphics{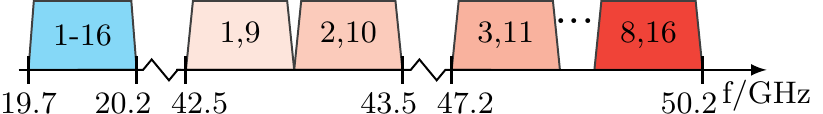}
	\caption{Single polarization frequency plan for the up- and the downlink: \gls{ffr} is applied in the downlink (Ka-band), so all 16 beams share a common frequency band. The multiplexing gain enables a parallel transmission of two feed signals by the feeder link within the same frequency segment of the V band. The numbers indicate the corresponding beam according to Fig.\ \ref{fig:HTS_SystArch}}.
	\label{fig:freqPlanMimosiso}
\end{figure}

The \gls{mimo} \gls{hts} system architecture is depicted in Fig. \ref{fig:HTS_SystArch}. Here we concentrate on the forward link, i.e. the link from the gateway to the users. 
Since a complete transmission chain with the uplink and the downlink is now considered, the mathematical notation to distinguish both parts will be adapted accordingly.

\subsubsection{\gls{mimo} Feeder Uplink}

The Earth portion of the \gls{mimo} feeder link consists of \mbox{$\AntNumTx=2$} gateway antennas, separated by several kilometers (\SIrange{30}{50}{\kilo\meter}) and inter-connected via a central processing unit which supervises the generation of the transmit signals. 
Time and phase synchronization of the antennas is ensured. \Acrshort{rf}-over-fiber transport can, for example, be used for this purpose %
\cite{Lau2014}. This technology has been successfully applied at the \acrshort{nasa} deep space network, where antennas separated by more than \SI{10}{\kilo\meter} must be synchronized for deep space communications. 
In the following, it is assumed that the inter-connection between the gateway antennas is perfect.

At the satellite, $\AntNumSLrx=2$ receive antennas are positioned in the orbital plane $\antspacingsl=\SI{3}{\meter}$ apart, constituting the receive array of the \gls{mimo} feeder link. To operate this feeder link, the bands \SIrange[range-phrase=-,range-units=single]{42.5}{43.5}{\giga\hertz} and \SIrange[range-phrase=-,range-units=single]{47.2}{50.2}{\giga\hertz} are exploited such that 
a total bandwidth of \SI{4}{\giga\hertz} per polarization state is available. The receive antennas cover the same geographical region centered in the middle of the gateway array, and the assumption is made that their beamwidth is sufficiently large. In that way, the antennas are relatively close to the beam center, and the receive antenna gain is maximum for all transmit-receive antenna links.

The satellite payload relies on a bent-pipe architecture with cross-strapping from the V-band feeder uplink to the Ka-band user downlink. No signal processing is considered in the payload to keep the satellite complexity low. Nevertheless, we observe that some sample-based signal processing could be envisioned.  
An architecture supporting digital transparent processing, known as \emph{digital bent-pipe}, would then be required. We will leave the further development of this possibility for future research. Some hints can be found also in \cite{Angeletti2008,Sulli2017}. 

It has to be noted that, to support the large aggregate user link bandwidth of a \gls{hts} system, several tens of spatially separated feeder links with a full reuse of the available frequency band are actually required in practice \cite{Vidal2012a}. To this end, several feeds are installed on the receive reflectors in order to cover different sites on Earth. In this case, an advantage of the \gls{mimo} approach lies in the fact that the necessary number of spatially separated links can be halved compared to state-of-the-art \gls{siso} links (e.g. 15 \gls{mimo} feeder links instead of 30 \gls{siso} links) \cite{Delamotte2018a}. Meanwhile, the total number of active gateway antennas in the system remains the same since two gateway antennas per feeder link are now used. 

The resort to \gls{mimo} feeder links can be intuitively seen as a solution to rearrange the antennas such that part of the interference between typical \gls{siso} feeder links is transformed into an information-bearing signal using spatial multiplexing. With a \gls{mimo}-based architecture, the angular separation of the different feeder links in a given region of deployment (e.g. Northern America) can indeed be increased to improve the beam isolation. In general, a distance of several hundreds of kilometers is required between the sites to guarantee sufficiently high \glspl{cir}. 

Here, the uplink part of the considered forward link consists of a single feeder link because the goal of the example is simply to illustrate how spatial multiplexing can be realized in a given link. The downlink of the studied \gls{hts} system will be dimensioned such that its sum user link bandwidth equals the sum bandwidth that can be supported by its uplink.    

\subsubsection{Multibeam Downlink}

The downlink part of the \gls{hts} system is made of $\AntNumSLtx=16$ Ka-band user beams delivering different data contents to fixed single antenna \acrfullpl{ut}. On the satellite a \gls{sfpb} architecture with $\AntNumSLrefl=4$ multibeam reflectors is considered. The reflectors are geometrically arranged as a uniform circular array with a diameter of \SI{3}{\meter}. 

On Earth a total of $\UserNumTot$ \glspl{ut} are uniformly distributed over the area covered by the $\AntNumSLtx$ beams. These \glspl{ut} are, for example, conventional single-antenna installations on a roof top of a building to serve households with individual data traffic. The comparison in the results section will be made to the conventional \gls{fr4} scheme where the same distribution of $\UserNumTot$ customers is assumed. This ensures a fair comparison between both schemes and allows a general conclusion from the simulation results. These results can be further extended to other user distributions, scheduling approaches or precoder designs from the literature, e.g. \cite{Icolari2017, Qian2014}. 

A downlink bandwidth of \SI{500}{\mega\hertz} within the range of \SIrange[range-phrase=-,range-units=single]{19.7}{20.2}{\giga\hertz} is available for the entire service zone. Since an \gls{ffr} scheme is addressed, this frequency band is jointly utilized by all beams. 

It is known that the achievable multiplexing gain offered by \gls{mu-mimo} is limited by the number of channel inputs. Only a group of up to $\AntNumSLtx=16$ \glspl{ut} can be served simultaneously via \gls{sdma}. Thus, the user downlink forms at most a $16\times 16$ \gls{mimo} channel. Since $\UserNumTot\gg\AntNumSLtx$, additional user scheduling is necessary to build groups of \glspl{ut} that will be served in different orthogonal \emph{resource blocks}. These different resource blocks can be, for example, separate time or frequency slots based on a \gls{tdma} or \gls{fdma} scheme, respectively. Our approach to schedule the $\UserNumTot$ users will be explained in more detail in Section \ref{subsec:UserScheduling}.    
\subsection{Channel Model}
\label{subsec:HTSchannelModel}

\subsubsection{\gls{mimo} Feeder Uplink}

The \gls{mimo} feeder link channel is modeled in the equivalent baseband using a block diagonal matrix  
\begin{align}
	\CTMblktilde =& \mydiag{\CTMultilde{1},\ldots,\CTMultilde{\frac{\AntNumSLtx}{2}}}~\in\mathds{C}^{\AntNumSLtx \times \AntNumSLtx},
	\label{eq:MIMOfeederUplink}
\end{align}
where $\CTMultilde{\countf}$ is a $2\times 2$ channel matrix between the $\AntNumTx=2$ gateway antennas and the $\AntNumSLrx=2$ satellite receive antennas. The index $\countf$ distinguishes the $\AntNumSLtx/2$ different center frequencies required due to the use of \gls{fdma} in addition to \gls{sdma} in the feeder uplink. Each entry $\CTMentryultilde{\countf,z}{n}$ corresponds to the channel coefficient from the $\countNumTx$-th gateway antenna to the $\countNumSL$-th satellite receive antenna at the $\countf$-th center frequency and takes both the free space propagation and the atmospheric impairments into account. 

Similarly to (\ref{eq:ImpairedCTM}), the matrix $\CTMultilde{\countf}$ can accordingly be expressed as 
\begin{align}
  \CTMultilde{\countf} = \CTMul{\countf}\cdot\AtmMatrix \odot  \AntGainMatrixul,
		\label{eq:CTM}
\end{align}
with $\CTMul{\countf} \in \mathds{C}^{2 \times 2}$, the \gls{mimo} \gls{los} channel matrix whose free space propagation coefficients are determined according to (\ref{eq:loschannelcoefficient}). Again, the matrix $\AtmMatrix \in \mathds{C}^{2 \times 2}$ is a diagonal matrix modeling the atmospheric impairments experienced by the gateway antennas. We observe that, in contrast to (\ref{eq:ImpairedCTM}), the multiplication with the matrix $\AtmMatrix$ is performed on the right-hand side since an uplink channel is now considered. The atmospheric impairments for the $\countNumTx$-th earth station should thus affect the $\countNumTx$-th column of $\CTMultilde{\countf}$. 

As already mentioned, rain attenuation represents a severe impairment in the Q/V band and imposes strong constraints on the link budget. It has motivated the development of gateway diversity schemes to ensure system availability \cite{Gharanjik2013a,Kyrgiazos2014a,Jeannin2014,Gharanjik2015}. In the sequel, the assumption is made that rain attenuation is the only weather impairment affecting the receive power in the feeder link. Other types of fading effects can indeed be compensated by an uplink power control scheme. 
In the results section, the proposed \gls{mimo} feeder link will be evaluated for different rain attenuations $\AtmAtt{1}$ at the first gateway antenna given a fixed $\AtmAtt{2}$ at the second antenna. 

The elements of matrix $\AntGainMatrixul \in\mathds{C}^{2\times 2}$ model the normalized radiation patterns of the satellite receive antennas.  
They can be calculated as  
\cite{Caini1992}
\begin{align}
    \AntGainUL{\countNumSL}{\countNumTx} =   \mybessel{1}{\counteru_{\countNumSL\countNumTx}}/2\counteru_{\countNumSL\countNumTx} + 36\mybessel{3}{\counteru_{\countNumSL\countNumTx}}/\counteru_{\countNumSL\countNumTx}^3, 
		\label{eqn:antenna_pattern}        
\end{align}
with $\counteru_{\countNumSL\countNumTx}=\pi\diameter/\carrierwavelength \sin(\offaxisangle{\countNumSL}{\countNumTx})$, and $\mybessel{1}{\counteru_{\countNumSL\countNumTx}}$ and $\mybessel{3}{\counteru_{\countNumSL\countNumTx}}$ being the Bessel functions of first kind and order one and three, respectively. We assume the same diameter $\diameter$ for all satellite receive antennas, and $\offaxisangle{\countNumSL}{\countNumTx}$ is the off-axis angle of the $\countNumSL$-th beam's boresight to gateway antenna $\countNumTx$.

In order to take the best advantage of spatial multiplexing in the feeder link, the \gls{los} uplink channel matrix $\CTMul{\countf}$ will be designed according to the criterion (\ref{eq:FinalOptCondition}). More precisely, the antenna geometry in the feeder link will be optimized such that a scaled unitary matrix $\CTMul{\countf}$ is obtained. It will be shown \deleted[id=td]{in the results section }that, in this case, the sum achievable rate of the \gls{hts} system is maximized.

\subsubsection{Multibeam Downlink}
\newcommand{\Transpose}[1]{{#1}^{\Tr}}
\newcommand{\Hermitian}[1]{{#1}^{\He}}
\newcommand{\CTMdlffr}{\tilde{\mymatrix{H}}_\text{d}}
In case of \gls{ffr}, an arbitrary \gls{ut} can potentially receive signal portions from all feeds. Therefore, the downlink channel $\CTMdlffr$ between the $\AntNumSLtx$ feeds and a group of $\UserNum$ \glspl{ut} is modeled as a densely populated matrix 
\begin{equation}
    \CTMdlffr = \CTMdl\odot\AntGainMatrixdl~\in\mathds{C}^{\UserNum\times\AntNumSLtx}.
\end{equation}
Here, $\CTMdl \in\mathds{C}^{\UserNum\times\AntNumSLtx}$ denotes the channel matrix that models the free space propagation according to (\ref{eq:loschannelcoefficient}) between the $\AntNumSLtx$ feeds and the $\UserNum$ households, which are simultaneously served with data in one recourse block. %
Equivalent to the uplink, the matrix $\AntGainMatrixdl \in\mathds{C}^{\UserNum\times\AntNumSLtx}$ describes the radiation patterns of the downlink multibeam antennas. The element $\matrixentry{\AntGainMatrixdl}{\user}{\countNumSL}=\AntGainDL{\user}{\countNumSL}$ denotes the normalized antenna gain from beam $\countNumSL$ to user $\user$. It is calculated using (\ref{eqn:antenna_pattern}) by an appropriate choice of the antenna parameters.

Similar to the feeder link channel, we are seeking to design an optimal downlink channel matrix $\CTMdlffr$, in which all row vectors are pairwise orthogonal. In this case the channel capacity is maximized according to (\ref{eq:capacitymax}). In contrast to the feeder link, the locations of the user antennas are arbitrary and an analytical condition as presented in (\ref{eq:FinalOptCondition}) cannot be applied here to derive optimal downlink matrices $\CTMdlffr$. To this end, a novel user grouping algorithm has been developed in \cite{Storek2017}, which addresses in particular the construction of downlink matrices $\CTMdlffr$ with pairwise orthogonal row vectors. We will apply and extend this algorithm to our \gls{hts} system proposal. It will be briefly recapped in Section\ \ref{subsec:UserScheduling}. The major difference of our approach compared to the state-of-the-art is reflected in the fact, that the condition of pairwise orthogonality inherently uses the signal phase as a design criterion whereas the great bunch of published work neglects the signal phase (see for example \cite{Guidotti2017} and references therein).

\newcommand{\AntGainMatrixSparse}{\tilde{\AntGainMatrix}}
\newcommand{\AntGainVecSparse}[1]{\tilde{\AntGainVec{#1}}}

\subsection{\gls{mimo} \gls{hts} System Model}
\label{subsec:SystemModel}
The equivalent baseband model of the \gls{hts} system under study is now introduced. 
Imperfections such as the non-linearities of the power amplifiers, phase noise, or differential delay and phase among multiple pathways in the satellite payload are assumed to be perfectly compensated using a calibration method \cite{Addio2010}. This correction guarantees that the downlink signals are phase-coherent to enable the feasibility of \gls{mu-mimo} precoding. 

Please note that, to enable the pre-processing of the transmit signals in the gateways, \gls{csi} about the uplink and downlink \gls{mimo} channels is necessary.  
To obtain this \gls{csi}, existing channel sounding strategies can be applied to estimate the amplitude and phase of the channel matrices $\CTMultilde{\countf}$ and $\CTMdlffr$. One solution consists, for example, in the transmission of orthogonal training sequences like the \gls{cazac} sequence \cite{Benvenuto2002} in the forward link. Applying the method as proposed in \cite{Hofmann2016}, the phase and amplitude information of the channel coefficients in $\CTMultilde{\countf}$ and $\CTMdlffr$ can be estimated via a cyclic cross-correlation of the signal received by the \glspl{ut} with the known sequence. This information can be fed back to the gateways via the return link to enable the pre-processing of the transmit signals. A small part of the available resources in the system must, therefore, be reserved for the transmission of a training sequence and the estimation of the channel. However, this is not specific to the approach presented here but applies to all types of precoding schemes for multibeam satellites.

We assume ideal \gls{csi} about the uplink and downlink \gls{mimo} channels in the sequel. It is known that the system performance degrades if the \gls{csi} is imperfect or outdated. The performance of different precoding schemes with imperfect \gls{csi} has been investigated in the literature, e.g. in \cite{Christopoulos2014,Wang2018}, and the interested reader is kindly referred to these papers and the references therein. 

\begin{figure*}[t]
	\centering
	\footnotesize
	\includegraphics{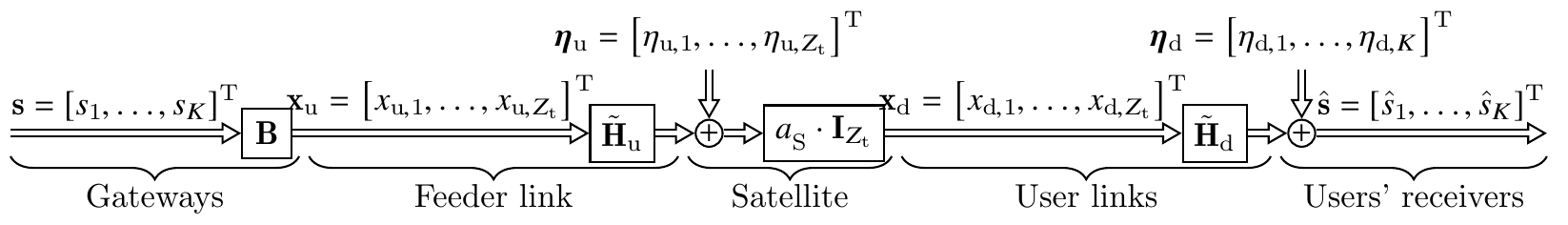}
	\caption{Block diagram of the transmission chain}
	\label{fig:MU_MIMOrelayingBlock}	
\end{figure*}

Fig. \ref{fig:MU_MIMOrelayingBlock} shows a block diagram of the transmission chain with the associated notations. For the sake of a compact notation, time indices are neglected in the sequel.  
Let 
\begin{equation}
	\symboltxvec=\left[\symboltxvecentry{1},\ldots,\symboltxvecentry{\UserNum}\right]^{\Tr} \in \mathds{C}^{\UserNum \times 1}
\end{equation}
be the vector of data symbols to be transmitted in a given time slot to a group of $\UserNum$ non-cooperative single-antenna \glspl{ut}, where $\symboltxvecentry{\user}$ is the symbol for the $\user$-th user. Please note that a pure unicast scenario is considered and that, therefore, $\symboltxvecentry{\user}$ can be different for different $\user$. These symbols are chosen from a constellation alphabet\footnote{Modulations are chosen according to the \acrshort{dvb-s2x} standard \cite{EuropeanTelecommunicationsStandardsInstitute2014}.} $\constellation$ with unit variance, and are uncorrelated such that $\myexpect{\symboltxvec\symboltxvec^{\He}} = \myeye{\UserNum}$. 

In the central processing unit of the gateway, a linear transformation of the data vector $\symboltxvec$ is performed using a precoding matrix $\precodingmatrix \in \mathds{C}^{\AntNumSLtx \times \UserNum}$. This matrix aims to mitigate the interference between the symbols transmitted in the same frequency channel in the uplink as well as in the downlink. %
Moreover, denoting $\eirpul$ the maximum \gls{eirp} per gateway antenna, the following condition must be fulfilled
\begin{equation}
\mytrace{\precodingmatrix^{\He}\qmatrixblk{\countNumTx}\precodingmatrix} \leq \eirpul\;,\quad \countNumTx=1,2. 
	\label{eq:PowerConstraintInGW}
\end{equation} 
The matrix $\qmatrixblk{\countNumTx}=\mydiag{\qmatrix{\countNumTx},\ldots,\qmatrix{\countNumTx}}$ is a $\AntNumSLtx \times \AntNumSLtx$ block diagonal matrix where $\qmatrix{\countNumTx}$ is a $2 \times 2$ matrix containing zeros in all entries except for the $\countNumTx$-th diagonal element which is equal to $1$. That is, in $\qmatrixblk{1}$ all odd and in $\qmatrixblk{2}$ all even diagonal entries are 1 while the remaining entries are zero. Condition (\ref{eq:PowerConstraintInGW}) corresponds to a per-antenna power constraint. It is required since the gateway antennas are equipped with their own \gls{hpa}. 

After the transmission over the \gls{mimo} feeder link, the symbols received by the satellite are 
\begin{equation}
	\signalrxvecul = \left[\signalrxveculentry{1},\ldots,\signalrxveculentry{\AntNumSLtx}\right]^{\Tr} = \CTMblktilde \precodingmatrix \symboltxvec  + \noisevecul. 
\end{equation}
The vector $\noisevecul=\left[ \noisevecentryul{1},\ldots,\noisevecentryul{\AntNumSLtx} \right]^{\Tr}$ is the vector of uplink circularly-symmetric complex Gaussian noise and is uncorrelated with the data symbols. 
In the sequel, the variance per real dimension of the complex uplink noise process is equal for all $\AntNumSLtx$ receive branches and is denoted by $\noisepowul{}$.

In the satellite payload, the coefficient $\amplgainsl{}$ models the amplification of the \glspl{hpa}. 
The vector of channel inputs to the downlink is then given by
\begin{equation}
	\signaltxvecdl = \left[\signaltxvecdlentry{1},\ldots,			\signaltxvecdlentry{\AntNumSLtx}\right]^{\Tr} = 
	\amplgainsl{}\cdot\CTMblktilde \precodingmatrix \symboltxvec + \amplgainsl{}\cdot\noisevecul
\end{equation}      
with $\signaltxvecdlentry{\countNumSL}$ being the downlink signal transmitted by the $\countNumSL$-th feed. The downlink \gls{eirp} in each beam should not exceed a maximum value $\eirpdl$ such that 
\begin{equation}
	\matrixentry{\CovMattxsigdl}{\countNumSL}{\countNumSL}\leq\eirpdl, ~\countNumSL\in\left\{1,\ldots,\AntNumSLtx\right\},
\label{eq:PowerConstraintInSat}
\end{equation}
where $\CovMattxsigdl = \myexpect{\signaltxvecdl\signaltxvecdl^{\He}}$ is the autocorrelation matrix of $\signaltxvecdl$. 
The gain $\amplgainsl{}$ is consequently chosen to fulfill the constraint (\ref{eq:PowerConstraintInSat}) with equality for at least one of the feeds. In other words, at least one downlink beam provides the maximum downlink \gls{eirp} $\eirpdl$ while the \gls{eirp} of the remaining beams can be lower or, in the best case, equal to $\eirpdl$. 

We note that the modeling of the satellite through a simple amplification coefficient follows from the assumption of an analog transparent architecture. In the case of a \emph{digital bent-pipe}, a sample-based processing of the signals could be envisioned. 
The satellite payload would then be modelled by a non-diagonal relaying matrix $\CTMslblk\in\mathds{C}^{\AntNumSLtx \times \AntNumSLtx}$. 

In the user links, the receive symbols are 
\begin{equation}
	\symbolrxvec = \left[\symbolrxvecentry{1},\ldots,\symbolrxvecentry{\UserNum}\right]^{\Tr} = \amplgainsl{}\cdot\CTMdlmod\CTMblktilde\precodingmatrix \symboltxvec + \amplgainsl{}\cdot\CTMdlmod\noisevecul + \noisevecdl,
\label{eq:RxDataVec}
\end{equation}
where $\noisevecdl=\left[\noisevecentrydl{1},\ldots,\noisevecentrydl{\UserNum}\right]^{\Tr}$ is the vector of downlink circularly-symmetric complex Gaussian noise. This noise is uncorrelated with both the data symbols $\symboltxvec$ and the uplink noise $\noisevecul$. Assuming similar receiving equipment for all households, the variance per real dimension of the noise process at each \gls{ut} is identical and represented by $\noisepowdl{}$. 

We define the receive \gls{cnr} at the \added[id=rs]{beam center as}\deleted[id=rs]{center of a spot beam as}
\begin{align}
    \CnrAtBeamCenter = {\FSLdl{}^2 \cdot \eirpdl}/\left( {2\noisepowdl{}} \right),
\end{align}
where $\FSLdl{}$ is the free space downlink gain as defined in (\ref{eq:loschannelcoefficient}).   

\subsection{User Scheduling}
\label{subsec:UserScheduling}
As stated before, only up to $\UserNum=\AntNumSLtx$ households can be simultaneously served within a single resource block. In order to supply all users with individual data, scheduling is necessary, i.e. the $\UserNumTot$ \glspl{ut} must be divided into groups of at most $\AntNumSLtx$ users. Hence, in addition to spatial multiplexing, a further multiple access scheme like \gls{tdma} \deleted[id=rs]{or \gls{fdma}} is necessary to support all users with individual data. Note that, without spatial multiplexing, $\UserNumTot$ resource blocks would be necessary whereas the necessary resource blocks reduces to $\UserNumTot/\UserNum$ with spatial multiplexing. 

As a major finding of Section \ref{sec:mimosatcombasics} it can be stated that the achievable data rate of a multiuser \gls{mimo} \gls{satcom} scenario crucially depends on the location of the involved antennas. Since the locations of the users are arbitrary and do not follow any regular placement, an analytic solution for the optimal placement, like the one that has been derived in (\ref{eq:FinalOptCondition}) for \gls{ula} arrangements, cannot be found here. However, the requirement of pairwise orthogonal channel vectors as formulated in (\ref{eq:ConditionOrthCTM}) still holds. 

Assume an optimal setup in the sense of (\ref{eq:ConditionOrthCTM}), then all $\UserNum$ \glspl{ut} of a group have orthogonal channel vectors. 
In this case, \gls{zf} precoding does not suffer from any power penalties due to the channel inversion, and can, therefore, achieve the \gls{dpc} rate region. However, the probability that such a set of \glspl{ut} exists is zero for a finite $\UserNumTot$ \cite{Swannack2004}. Therefore, we aim at combining \glspl{ut} with channel vectors that are ``nearly'' or ``at most'' orthogonal. The orthogonality between the $i$-th and the $j$-th \gls{ut} is quantified by means of the scalar product of the channel vectors
\newcommand{\Norm}[1]{\left\Vert #1 \right\Vert}
\begin{align}
	\cos\left(\measuredangle \left( \CTVdltilde{i}, \CTVdltilde{j} \right) \right) = { \myabs{ \Hermitian{\CTVdltilde{i}} \CTVdltilde{j}  } } / \left( {\Norm{\CTVdltilde{i}}\Norm{\CTVdltilde{j}}} \right), 
	\label{eqn:def_of_coc}
\end{align}
where $\measuredangle ( \CTVdltilde{i}, \CTVdltilde{j}) $ refers to the enclosed angle of the vectors $\CTVdltilde{i}$ and $\CTVdltilde{j}$. The channel vectors are orthogonal if $\cos(\measuredangle ( \CTVdltilde{i}, \CTVdltilde{j})) = 0$. 

In \cite{Storek2017} the 
Multiple Antenna Downlink Orthogonal Clustering (MADOC) scheduling algorithm has first been published, in which (\ref{eqn:def_of_coc}) is used as a metric to determine the spatial compatibility between two \glspl{ut} and to build up so-called $\epsilon$-orthogonal groups. The main idea is to assemble \glspl{ut} in a common group only if (\ref{eqn:def_of_coc}), calculated for all combinations of \glspl{ut} within a group, does not exceed a certain threshold $\epsilon$. The proposed algorithm in \cite{Storek2017} is used in this paper to schedule all households into $\numOfGroups$ disjoint groups. In the optimal case, the downlink channel matrix $\CTMdlffr$ is then composed of perfectly orthogonal row vectors, and each household in each group receives then individual data content without any \gls{cci}. 
While we utilize the MADOC algorithm in this paper to form user groups for unicast transmission over several beams, its metric can similarly be applied to realize a multicast transmission over a single beam. In this case, (\ref{eqn:def_of_coc}) aims to identify channel vectors that are parallel.
For the first time, this algorithm takes advantage of the spatial orthogonality offered by the \gls{mimo} channel, and the spatial dimension is exploited as a further degree of freedom.

This is indeed a great advantage of the \gls{mimo} technology. In other words, although $\UserNum$ users still occupy the same frequency band and polarization like it would be the case in a broadcast application, their spatial separation is now exploited to distribute individual content to different households. This way, MIMO answers the user's expectations for individual content like \gls{vod} and supports the transition from broadcast to unicast transmission in future satellite implementations.

\newcommand{\totalUserSet}{\bar{\mathcal{K}}}
\newcommand{\selectedUserSet}{\mathcal{K}}
\newcommand{\groupLetter}{\mathcal{T}}
\newcommand{\group}[1]{\groupLetter^{(#1)}}
\newcommand{\groupingThreshold}{\epsilon}

\subsection{Precoding Design}
\label{subsec:PrecoderDesign}
In our \gls{hts} scenario, maximal fairness between the selected users should be ensured by the precoder. In a unicast scenario, every customer indeed expects a minimum assured data rate that is guaranteed by the satellite provider according to the contract model. This is fundamentally different to the broadcast scenario where all users receive the same data and the quality-of-experience solely depends on the individual receiving equipment. 

The precoding matrix $\precodingmatrix$ will be optimized using the common \gls{zf} criterion. It is known to offer close-to-optimal performance when the noise contribution is limited, which is the case for feeder links, or for a large number of channel outputs as it is the case in the user links \cite{Caire2003}. Here, the following condition must be fulfilled:
\begin{equation}
    \CTMdltilde\CTMblktilde\precodingmatrix = \mydiag{\ScalingPrecVec}\;,
		\label{eq:JointZFcond}
\end{equation}
with $\ScalingPrecVec=\left[\ScalingPrec_1,\ldots,\ScalingPrec_{\AntNumSLtx}\right]^{\Tr}$, a vector of non-negative numbers. A max-min fairness optimization problem can then be formulated as
\newcommand{\MatrixElement}[3]{  {\left[ {#1} \right]}_{#2,#3} }
\begin{align}
	\max_{\ScalingPrecVec\ge \myvector{0},\precodingmatrix,\amplgainsl{}} \hspace{10pt} 
	& \min_\user \amplgainsl{}\ScalingPrec_{\user} \nonumber \\
	\text{s.t.} \hspace{10pt}                              
	& \CTMdltilde\CTMblktilde\precodingmatrix = \mydiag{\ScalingPrecVec} \nonumber \\
& \mytrace{\precodingmatrix^{\He}\qmatrixblk{\countNumTx}\precodingmatrix} \leq \eirpul\;,\quad \countNumTx=1,2	                                                                        \nonumber \\
	                                                       &  \amplgainsl{2}\cdot\MatrixElement{\CTMblktilde\precodingmatrix \precodingmatrix^{\He} \CTMblktilde^{\He}+2\noisepowul{}\myeye{\AntNumSLtx}}{\countNumSL}{\countNumSL} \le \eirpdl\;, \forall\;\countNumSL.
   	\label{eqn:calc_fair_precodM_joint}
\end{align}
The solution of (\ref{eqn:calc_fair_precodM_joint}) entails $\ScalingPrecVec=\ScalingPrec{}\myvector{1}$ \cite{Wiesel2008}. Moreover, since the noise contribution in the feeder link is limited, \deleted[id=rs]{Moreover, the noise contribution in the uplink is limited since feeder links are designed with a proper link budget. Hence,} the third constraint of (\ref{eqn:calc_fair_precodM_joint}) can be relaxed by assuming that $2\noisepowul{}\ll\MatrixElement{\CTMblktilde\precodingmatrix \precodingmatrix^{\He} \CTMblktilde^{\He}}{\countNumSL}{\countNumSL}$, $\forall \countNumSL$, and the optimization problem is now expressed as 
\begin{align}  
		\max_{\ScalingPrec\ge 0,\mathbf{W},\amplgainsl{}} \hspace{10pt} & \amplgainsl{}\ScalingPrec \nonumber \\
		\text{s.t.} \hspace{10pt}                              & \ScalingPrec^2\cdot\mytrace{\bar{\precodingmatrix}^{\He}\qmatrixblk{\countNumTx}\bar{\precodingmatrix}} \leq \eirpul\;,\quad \countNumTx=1,2 \nonumber \\
		& \amplgainsl{2}\ScalingPrec^2\cdot\MatrixElement{\CTMblktilde\bar{\precodingmatrix} \bar{\precodingmatrix}^{\He} \CTMblktilde^{\He}}{\countNumSL}{\countNumSL} \le \eirpdl\;, \forall\;\countNumSL, 
		\label{eqn:calc_fair_precodM_jointReformulated}
\end{align}
where $\bar{\precodingmatrix}=\CTMblktilde^{\mpi}\CTMdltilde^{\mpi}+\mathbf{P}_{\perp}\mathbf{W}$ and $\mathbf{P}_{\perp}$ is the orthogonal projection into the null space of $\CTMdltilde\CTMblktilde$. Matrix $\mathbf{W}$ is an arbitrary complex matrix. The reformulation of the optimization problem relies on the fact that a precoding matrix fulfilling the condition of (\ref{eq:JointZFcond}) is of the form $\precodingmatrix=\ScalingPrec\bar{\precodingmatrix}$ \cite{Wiesel2008}. It corresponds to a generalized inverse of $\CTMdltilde\CTMblktilde$. A solution to (\ref{eqn:calc_fair_precodM_jointReformulated}) is then obtained with
\begin{align}
	\ScalingPrec &= \sqrt{\eirpul/\mymaxvalue{\mytrace{\bar{\precodingmatrix}^{\He}\qmatrixblk{\countNumTx}\bar{\precodingmatrix}}}}\;,\\ 
	\amplgainsl{} &= \sqrt{\eirpdl/\ScalingPrec^2\cdot\mymaxvalue{\MatrixElement{\CTMblktilde\bar{\precodingmatrix} \bar{\precodingmatrix}^{\He} \CTMblktilde^{\He}}{\countNumSL}{\countNumSL} }}\;,
\end{align}
and the matrix $\bar{\precodingmatrix}$ is determined with the following second-order cone program
\begin{equation} 
	\min_{\mathbf{W},t} t\quad \text{s.t.} \quad \MatrixElement{\CTMblktilde\bar{\precodingmatrix} \bar{\precodingmatrix}^{\He} \CTMblktilde^{\He}}{\countNumSL}{\countNumSL} \leq t\;. 
\end{equation}
Interestingly, the precoding matrix $\precodingmatrix$ obtained with (\ref{eqn:calc_fair_precodM_jointReformulated}) leads to the same performance as the precoder $\precodingmatrix = \ScalingPrec \precodingmatrixul \precodingmatrixdl$ where $\precodingmatrixul$ and $\precodingmatrixdl$ are the solutions of a max-min fairness optimization problem for the downlink and the uplink, respectively.\footnote{%
Note that, in this case, \gls{csi} of $\CTMblktilde$ and $\CTMdltilde$ must be available instead of the entire link from the gateway to the \glspl{ut}. \added{To obtain $\CTMblktilde$ and $\CTMdltilde$ separately, ad-hoc and different training sequences for the feeder link and the user link are required.} \deleted{This can be achieved if the channel sounding method as proposed in \mbox{\cite{Hofmann2016}} is applied. 
Following the approach in \mbox{\cite{Hofmann2016}}, a single user is able to estimate the relative phase and amplitude information of the feeder uplink $\CTMblktilde$. Feeding back this information to the gateway, the relative phase and amplitude information of $\CTMdlffr$ can be estimated in the central processing unit by comparing the estimation results of each \gls{ut}.} %
} %
The generalized inverse of the product of $\CTMdltilde$ with the square non-singular matrix $\CTMblktilde$ is indeed equal to the product of the inverse of $\CTMblktilde$ and the generalized inverse of $\CTMdltilde$ \cite{Lutkepohl1996}. Simulation results in Section\ \ref{subsec:SimulationResults} will validate the equality of both approaches. \added{A cascade design takes into account the fact that the matrices $\precodingmatrixul$ and $\precodingmatrixdl$ do not need to be updated at the same rate. In a practical system, the matrix $\CTMdltilde$ and, thus, the precoder $\precodingmatrixdl$ will change at a rate of a few milliseconds. This fast change is not due to the coherence time of the downlink channel but to the user scheduling. Different groups are indeed served in successive time slots. On the other hand, the characteristics of the uplink \gls{mimo} channel $\CTMblktilde$ do not evolve so fast, as the same fixed gateway antennas are always used. The only source of random variation in the uplink are the weather impairments. Their coherence time is at least a few hundreds of milliseconds, and the same uplink precoding matrix $\precodingmatrixul$ can be used during this time.}

\subsection{Comparison to the State-of-the-Art}
\label{subsec:CompStateOfTheArt}
For the first time we address both, the application of spatial \gls{mimo} in the feeder link and the use of \gls{ffr} in the user link over a transparent \gls{geo} relay. 
System architectures in the existing literature are limited to  either \gls{mimo} in the feeder link\footnote{%
To the best of our knowledge, the two paper \cite{Delamotte2018a,Delamotte2018b} are the only contributions so far considering the concept of \gls{mimo} in the feeder links.} %
or \gls{mimo} related approaches based on \gls{ffr} with a multibeam architecture in the downlink (e.g. in \cite{Arnau2012,Vazquez2016,Wang2018}).

The second fundamental difference to the current state-of-the-art is that in our approach the signal phase of the \gls{los} part of the channel is exploited. 
The vast majority of publications do not consider the signal phase of the \gls{los} satellite channel in their models. For example, the authors in \cite{Joroughi2017,Christopoulos2015,Diaz2007} assume a uniformly distributed channel phase in the downlink, irrespective of the user's location and the position of the feed on the payload. The authors in \cite{Guidotti2017}\deleted{even neglect the phase information} \added{state that the phase information in the design of the precoder is neglected}. 
Therefore, the grouping of users \deleted[id=rs]{for multicast or unicast applications}as well as the precoder design can only rely on the amplitude information. 
Simulation results in Section\ \ref{subsec:SimulationResults} will show that, in this case, the achievable rate is limited. A payload with multiple spatially separated antennas, as proposed here, would have no advantage compared to a payload with a single reflector if only the amplitude information is exploited.

While neglecting the phase of the \gls{los} path might be valid as long as a single antenna on the satellite is assumed, it becomes inappropriate if multiple antennas are considered. Many satellites currently under procurement offer at least four reflectors mounted as side-deployable antennas at a separation of \SIrange{6}{10}{m}. These architectures offer the chance to exhibit additional \deleted[id=rs]{spatial }multiplexing gains when resorting to schemes that take the signal phase\deleted[id=rs]{ information} into account.

\subsection{Performance Criterion}
\newcommand{\CTMtot}{\mymatrix{C}} 
To assess the performance of the proposed \gls{mimo} \gls{hts} system the sum achievable rate is used. The sum achievable rate can be obtained by first computing for each user its input-output mutual information. To this end, 
we define an auxiliary model from (\ref{eq:RxDataVec}). 
Using $\CTMtot=\amplgainsl{}\CTMdlmod\CTMblktilde\precodingmatrix$, we can write 
\begin{align}
    \symbolrxvec &= \mydiag{\CTMtot}\symboltxvec + \left(\CTMtot-\mydiag{\CTMtot}\right)\symboltxvec + \amplgainsl{}\CTMdlmod\noisevecul + \noisevecdl \nonumber\\
                 &= \mydiag{\CTMtot}\symboltxvec + \noiseveceq,
    \label{eq:AuxiliaryChMod}             
\end{align}
where $\noiseveceq$ is a vector gathering the contributions of the uplink and downlink noise as well as the inter-stream interference if 
$\CTMtot$ is non-diagonal. We note that \deleted[id=rs]{the }matrix $\CTMtot$ will not be diagonal if $\CTMblktilde$ or $\CTMdlmod$ are rank deficient, i.e. that no \acrlong{zf} is possible. 

In practice a satellite link will be designed to avoid such a configuration. In the uplink this is ensured if the gateway antennas are located following the design criteria of (\ref{eq:FinalOptCondition}). In the downlink this is ensured if the users who are jointly considered have nearly\deleted[id=rs]{ or at most} orthogonal channel vectors according to (\ref{eq:ConditionOrthCTM}).   

Assuming a circularly-symmetric complex Gaussian distribution for the inter-stream interference, the channel law for the $\user$-th user in a given resource block is 
\begin{equation}
	\myprob{\symbolrxvecentry{k}|\symboltxvecentry{k}} = \frac{1}{2\pi\noisepoweq{,\user}} \cdot \exp\left\{ - \frac{\myabs{\symbolrxvecentry{k}-\CTMentrytot{\user}{\user}\symboltxvecentry{k}}^2}{2\pi\noisepoweq{,\user}} \right\}
	\label{eq:ChannelLawUser}
\end{equation}
with $\CTMentrytot{\user}{\user}$ the $\user$-th diagonal element of $\CTMtot$ and $2\noisepoweq{,\user}$ the variance of the $\user$-th entry of the \deleted[id=rs]{complex }vector $\noiseveceq$.

With the knowledge of (\ref{eq:ChannelLawUser}), the input-output mutual information in \si{\bit\per\channeluse} for the $\user$-th user is given by 
\begin{equation}
    \miuser{\user} = \myexpect{\mylog{2}{\frac{\sum_{\symboltxvecentry{}\in\constellation}\myprob{\symbolrxvecentry{k}|\symboltxvecentry{}}}{\myprob{\symbolrxvecentry{k}|\symboltxvecentry{k}}}}}.\label{eq:MIk}
\end{equation}
The application of (\ref{eq:MIk}) assumes a \gls{ml} symbol detection. 

Moreover, we introduce the spectral efficiency  
\begin{equation}
	\speceffuser{\user} = \log_{2}\left(1 + \CINRuser{\user}\right),
	\label{eq:SpeceffUser}
\end{equation}
of the $\user$-th user channel for comparison, where $\CINRuser{\user}= \CTMentrytot{\user}{\user}/\noisepoweq{,\user}$ denotes the \gls{cinr} of user $\user$. %

According to the considered input constellation (discrete modulation alphabet or Gaussian distributed input symbols), the achievable rate in \si{\bit\per\second} is obtained from the mutual information $\miuser{\user}$ or the spectral efficiency $\speceffuser{\user}$ by normalization with the symbol duration $\symbolperiod$, i.e. $\TransRateUser{\user} = 1/\symbolperiod \cdot \miuser{\user}$ or $\TransRateUser{\user} = 1/\symbolperiod \cdot \speceffuser{\user}$. The sum achievable rate is then given by 
\begin{equation}
   \TransRate =  \frac{1}{\numOfGroups}\sum\limits_{\usergroupindex=1}^{\numOfGroups}\sum\limits_{\user=1}^{\UserNum^{\usergroupindex}} \TransRateUser{\user}^{\usergroupindex},
   \label{eqn:sumRate}
\end{equation}
where the superscript $\usergroupindex$ is introduced to take into account the fact that several user groups are considered. Thus, $\UserNum^{\usergroupindex}$ and $\TransRateUser{\user}^{\usergroupindex}$ denote the number of users and the achievable rate of user $\user$ in group $\usergroupindex$, respectively. 
Finally, the rate per user beam $\TransRatePerBeam$ is defined as $\TransRatePerBeam = \TransRate / \AntNumSLtx$.  %

The sum achievable rate $\TransRate$ defined in (\ref{eqn:sumRate}) together with the rate per user beam $\TransRatePerBeam$ allows us to assess the performance of the proposed \gls{mimo} \gls{hts} system. These two metrics are used in the next section to evaluate the benefit of spatial \gls{mimo} in the uplink and the downlink.

\section{\gls{mimo} \gls{hts} Simulation Results}
\label{subsec:SimulationResults}

The proposed \gls{mimo} uplink and downlink architectures will now be evaluated for a satellite positioned at the longitude $\lonsl = \SI{115}{\degree}\;\text{W}$. The feeder uplink operates in the V-band, and the user downlink is in the Ka-band. The gateway array has an orientation $\ulaorientes=\SI{0}{\degree}$. It is located at latitude $\lates= \SI{38}{\degree}\;\text{N}$ and longitude $\lones = \SI{98}{\degree}\;\text{W}$. Finally, the $\AntNumSLtx=16$ user beams in an \gls{sfpb} architecture cover the U.S. West Coast where a total of $\UserNumTot=\num{4000}$ households have to be served.

We assume a symbol rate of $1/\symbolperiod=\SI{10}{MHz}$ per customer. If the carrier spacing equals $1.05/\symbolperiod$, up to $\NumCarrier=\left\lfloor \SI{500}{MHz}\cdot\symbolperiod/1.05 \right\rfloor=47$ carriers via \gls{fdma} are supported by the \SI{500}{MHz} downlink beams.

To emphasize the benefits of spatial multiplexing for an \gls{hts} system, the uplink and the downlink of the system proposed in Section~\ref{sec:mimohtsexample} will be successively replaced by a state-of-the-art approach. Here, the two following configurations will be investigated in terms of the sum achievable data rate: 
\begin{itemize}
	\item A $2\times 2$ \gls{mimo} feeder link is compared to a \gls{siso} feeder link with single-site diversity, i.e. one \gls{siso} gateway antenna is active while the second \gls{siso} gateway antenna is in stand-by.  
	\item A \gls{mu-mimo} downlink is compared to a conventional \gls{fr4} \gls{siso} approach. 
\end{itemize}
In addition to the sum achievable data rate we also assess the vulnerability of the feeder link to rain fading events. As it has already been mentioned in Section \ref{subsec:ChannelImpairments}, a Q/V-band feeder link may suffer from rain fading events resulting in additional attenuations of several decibel. The simulations will show how \gls{mimo} feeder links can take advantage of an additional site diversity gain that directly increases the system availability.

\subsection{Feeder Link Performance}

\begin{table}[!t]
\footnotesize
\centering
\caption{Parameters for the feeder link evaluation} 
\label{ParamFeederLinkEval}
	\begin{IEEEeqnarraybox}[\IEEEeqnarraystrutmode\IEEEeqnarraystrutsizeadd{1pt}{0pt}]{u/c/x/c/x}
		\IEEEeqnarrayrulerow\\
		& \IEEEeqnarraymulticol{1}{t}{\bfseries\itshape \acrshort{siso} feeder link} && \IEEEeqnarraymulticol{1}{t}{\bfseries\itshape \acrshort{mimo} feeder link} \\
		\IEEEeqnarraystrutsize{0pt}{2pt}\\
		\IEEEeqnarraydblrulerowcut\\
		\IEEEeqnarraystrutsize{0pt}{1pt}\\
		 Frequency bands & \IEEEeqnarraymulticol{4}{t}{\SIrange[range-phrase=-,range-units=single]{42.5}{43.5}{\giga\hertz} + \SIrange[range-phrase=-,range-units=single]{47.2}{50.2}{\giga\hertz} 
		}\\
		 Downlink feeds $\AntNumSLtx$ & \IEEEeqnarraymulticol{4}{c}{ 		16} \\ 
		 \thead{Usable bandwidth per\\ downlink beam} & \SI{250}{\mega\hertz} && \SI{500}{\mega\hertz} \\
		 Gateway antenna gain & \IEEEeqnarraymulticol{4}{c}{\SI{61.4}{\dBi}} \\ 
		 \thead{Transmit power per \\ gateway antenna} & \SI{22}{dBW} && \SI{19}{dBW} \\
		 Satellite $\GoT$ & \IEEEeqnarraymulticol{4}{c}{\SI[per-mode=symbol]{26}{\deci\bel\per\kelvin}} \\ 
		\IEEEeqnarraystrutsize{0pt}{2pt}\\
		\IEEEeqnarraydblrulerowcut\\
			\IEEEeqnarraystrutsize{0pt}{1pt}\\
		 \itshape Downlink configuration & \\
		\IEEEeqnarraystrutsize{0pt}{1pt}\\											
		\IEEEeqnarrayrulerow\\ 
		\IEEEeqnarraystrutsize{0pt}{1pt}\\
		 Architecture & \IEEEeqnarraymulticol{4}{t}{\acrshort{mu-mimo} \acrshort{ffr}} \\
		 \thead{\acrshort{cnr} at the center \\ of a beam $\CnrAtBeamCenter$} & \IEEEeqnarraymulticol{4}{c}{\SI{10}{\deci\bel}} \\
		\IEEEeqnarraystrutsize{0pt}{2pt}\\
		\IEEEeqnarrayrulerow	
	\end{IEEEeqnarraybox}
\end{table}

The system parameters for the analysis of the feeder link are provided in Table\ \ref{ParamFeederLinkEval}. Here the objective is to 
assess the performance of a $2\times 2$ \gls{mimo} feeder link in terms of the sum achievable data rate. 
The available bandwidth in the feeder uplink is \SI{4}{GHz}. Due to the $2\times 2$ spatial multiplexing in the \gls{mimo} feeder link, a total of \SI{8}{GHz} usable bandwidth is available in the user downlink. $\AntNumSLtx=16$ downlink feeds should be supported and, thus, the households in each downlink beam can be served with \SI{500}{MHz} of bandwidth.\footnote{Please have in mind that we assume a transparent satellite payload.} The \gls{siso} feeder link instead can only support a bandwidth of \SI{4}{GHz} in total, and, therefore, the total bandwidth per downlink beam is set for this benchmark system to $\SI{250}{\mega\hertz}$. In this case, the number of carriers which can be supported via \gls{fdma} reduces to $\left\lfloor \SI{250}{MHz}\cdot\symbolperiod/1.05 \right\rfloor = 23$.

In order to still meet coordinated power density limits, the \gls{eirp} of the \gls{mimo} feeder link must be the same as of the \gls{siso} feeder link. Therefore, the transmit power per gateway antenna is \SI{3}{dB} less in the case of a \gls{mimo} gateway antenna. We assume a gateway antenna gain of \SI{61.4}{dBi}, and, together with a $\GoT$ of \SI{26}{\deci\bel\per\kelvin} per satellite antenna\footnote{The $\GoT$ of \SI{26}{dB/K} is based on an antenna with diameter of \SI{1.2}{m}, efficiency of 55\% at \SI{48}{GHz} and a system temperature of \SI{500}{K} \cite{Thompson2011}. Since the center of coverage of both satellite antennas is at the center of the gateway array, the offaxis angle from point of boresight is only \SI{0.017}{\degree}. The resulting depointing loss is only \SI{0.02}{dB}. This value is too small to make a difference in the simulation results and has,  therefore, been neglected.}, the feeder link budget is fully defined. Under clear sky conditions an uplink \gls{cnr} of \SI{24}{dB} is achieved.

For the user links a \gls{mu-mimo} \gls{ffr} approach with a \gls{cnr} of \SI{10}{\deci\bel} at the center of the spot beams is always assumed. It ensures that the differences in terms of sum achievable rate between the \gls{siso} and the \gls{mimo} feeder links are, in this case, only influenced by the uplink design.         

\begin{figure}[t]
	\centering
	\includegraphics{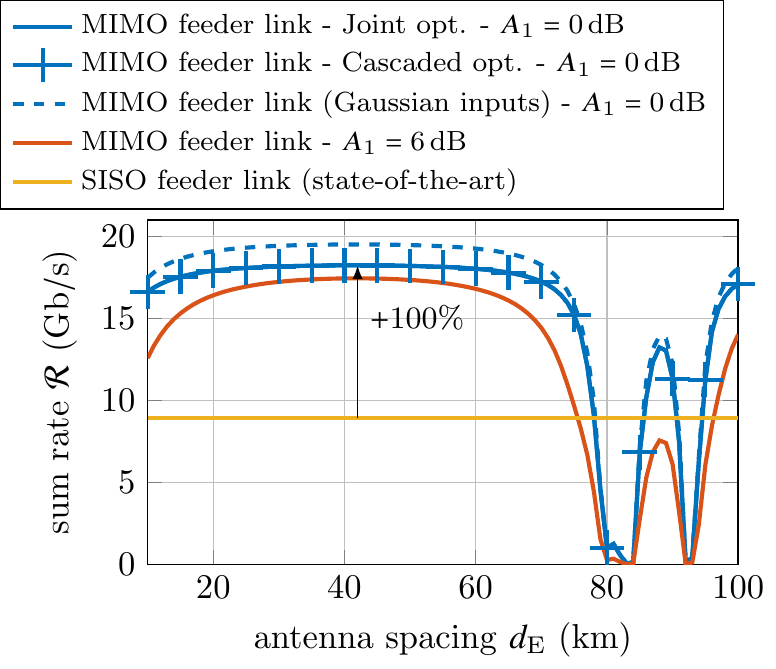}
	\caption{Sum rate vs. gateway antenna spacing $\antspacinges$ for two different weather conditions at gateway antenna one while gateway antenna two experience clear sky, i.e. $\AtmAtt{2}=\SI{0}{\deci\bel}$.} 
	\label{fig:FuncDistSumRate_FFR1600users10dB_UpMimo-Siso_Bu=500MHz-250MHZ_CtxMaxN4UL=251dB-254dB_CNRdl=10dB.tex}
\end{figure}

In Fig.\ \ref{fig:FuncDistSumRate_FFR1600users10dB_UpMimo-Siso_Bu=500MHz-250MHZ_CtxMaxN4UL=251dB-254dB_CNRdl=10dB.tex}, the influence of the gateway separation $\antspacinges$ on the sum achievable rate is shown for two different weather conditions. In one scenario, the first gateway antenna does not experience any rain attenuation, i.e. $\AtmAtt{1}=\SI{0}{\deci\bel}$, while the other scenario assumes a medium rain attenuation of $\AtmAtt{1}=\SI{6}{\deci\bel}$. The second gateway antenna benefits in both scenarios from clear-sky conditions, i.e. $\AtmAtt{2}=\SI{0}{\deci\bel}$. For the sake of illustration, results obtained with a joint optimization and a cascaded optimization of the precoding matrix are shown in the case $\AtmAtt{1}=\SI{0}{\deci\bel}$. The goal is to confirm that both optimizations lead actually to the same precoder. This property has been emphasized in Section~\ref{subsec:PrecoderDesign}. Moreover, for comparison purposes the sum achievable rate based on the spectral efficiency according to (\ref{eq:SpeceffUser}) is provided (dashed blue curve). Finally, the achievable rate of a state-of-the-art \gls{siso} feeder link is also displayed. In this configuration, the second gateway antenna is the only active ground station.  

The sum achievable rate is maximized for a separation of around \SI{40}{\kilo\meter} but a large range of \SI{\pm 15}{\kilo\meter} relative to the optimal position can be accepted without entailing serious performance degradations. The improved sum data rate of the \gls{mimo} solution is the result of spatial multiplexing in the feeder uplink. Although \SI{4}{GHz} of bandwidth is allocated in the feeder uplink, \SI{8}{\giga\hertz} of user link bandwidth instead of \SI{4}{\giga\hertz} for the state-of-the-art can be used to serve the \num{4000} households in the downlink. 
It can be noted that the minima for values of $\antspacinges$ near \SI{80}{\kilo\meter} and \SI{90}{\kilo\meter} are due to the presence of close-to-singular \gls{mimo} channel matrices in the bands \SIrange[range-phrase=-,range-units=single]{47.2}{50.2}{\giga\hertz} and \SIrange[range-phrase=-,range-units=single]{42.5}{43.5}{\giga\hertz}, respectively. However, distances of more than \SI{80}{\kilo\meter} do not belong to the region of interest for a practical system design. 

The curves show also that an additional atmospheric attenuation does not severely degrade the data rate performance of the \gls{mimo} feeder link. Even for a rain attenuation of $\AtmAtt{1}=\SI{6}{\deci\bel}$ at gateway antenna one, the \gls{mimo} feeder link still outperforms the state-of-the-art \gls{siso} approach even in clear-sky. One reason for this small degradation is clearly the \SI{14}{dB} higher uplink \gls{cnr} compared to the downlink \gls{cnr} (\SI{24}{dB} in the uplink vs. \SI{10}{dB} in the downlink). However, an additional attenuation of \SI{6}{dB} at one antenna does, nevertheless, not result in a \SI{6}{dB} lower \gls{cnr} in one of the two equivalent \gls{siso} sub-channels. In fact, since in an optimal \gls{mimo} channel both gateway antennas contribute equally to the receive \gls{cnr}, an attenuation of \SI{6}{dB} at one gateway antenna results only in a \SI{3}{dB} lower \gls{cnr} per receive antenna. Apart from the multiplexing gain, the additional spatial domain usage of the \gls{mimo} approach inherently provides a higher robustness against weather effects \cite{Knopp2010}. In other words: The results clearly suggest to avoid so-called ``cold redundancy'' \gls{wrt} gateway antennas, i.e. it is better to switch all available antennas on and operate them as a \gls{mimo} feeder link instead of preserving antennas for outage redundancy only.

The results presented in this section illustrate the potential of the spatial \gls{mimo} concept for the feeder links of \gls{hts} systems. The possibility to distribute spatially multiplexed signals containing individual content for multiple users perfectly supports the resort of the conventional \gls{tv} broadcast scenario to unicast transmission. Please note again that the goal of this simulation example was to illustrate how spatial multiplexing can generally be realized in a single feeder link. 

As already mentioned earlier, several tens of spatially separated feeder links are necessary in practice \cite{Vidal2012a}, and, in this case, the advantage of the \gls{mimo} approach lies in the reduction of the interference between neighboring feeder links. Less geographically separated \gls{mimo} feeder links than \gls{siso} feeder links are required to provide a given amount of data to serve all households with different content in a practical \gls{hts}. Therefore, if the feeder links have to be deployed in a certain country or continent, interference can be more easily avoided by guaranteeing a larger angular separation between the beams \cite{Delamotte2018a}. This is especially advantageous in terms of system availability as better link budgets reduce the probability of experiencing an outage when rain events affect the link quality \cite{Delamotte2018b}.

\subsection{User Link Performance}

\begin{table}[!t]
\footnotesize
\centering
\caption{Parameters for the user link evaluation} 
\label{ParamUserLinkEval}
	\begin{IEEEeqnarraybox}[\IEEEeqnarraystrutmode\IEEEeqnarraystrutsizeadd{1pt}{0pt}]{x/u/x/c/x/c/x}
		\IEEEeqnarrayrulerow\\
		&&& \IEEEeqnarraymulticol{1}{t}{\bfseries\itshape \acrshort{siso} \acrshort{fr4}} && \IEEEeqnarraymulticol{1}{t}{\bfseries\itshape \acrshort{mu-mimo} \acrshort{ffr}} \\
		\IEEEeqnarraystrutsize{0pt}{2pt}\\
		\IEEEeqnarraydblrulerowcut\\
		\IEEEeqnarraystrutsize{0pt}{1pt}\\
		& Downlink feeds $\AntNumSLtx$ && \IEEEeqnarraymulticol{4}{c}{16} \\ 
		& Bandwidth per beam && \SI{125}{\mega\hertz} && \SI{500}{\mega\hertz} \\
		& \thead{Downlink \gls{eirp} \\ per beam $\eirpdl$} &&  \IEEEeqnarraymulticol{4}{t}{\SIrange{51}{65}{\dBW}} \\ 
		& User terminal $\GoT$ && \IEEEeqnarraymulticol{4}{c}{\SI[per-mode=symbol]{16.9}{\deci\bel\per\kelvin}} \\
		& \thead{Resulting \acrshort{cnr} at \\ beam center $\CnrAtBeamCenter$} && \IEEEeqnarraymulticol{1}{t}{\SIrange{6}{20}{\deci\bel}} && \IEEEeqnarraymulticol{1}{t}{\SIrange{0}{14}{\deci\bel}} \\
		\IEEEeqnarraystrutsize{0pt}{2pt}\\
		\IEEEeqnarraydblrulerowcut\\
			\IEEEeqnarraystrutsize{0pt}{1pt}\\
		& \itshape Uplink configuration && \\
		\IEEEeqnarraystrutsize{0pt}{1pt}\\											
		\IEEEeqnarrayrulerow\\ 
		\IEEEeqnarraystrutsize{0pt}{1pt}\\
		& Architecture && \IEEEeqnarraymulticol{4}{t}{\acrshort{mimo} feeder link with $\antspacinges=\SI{40}{\kilo\meter}$} \\
		& \raisebox{-5pt}[0pt][0pt]{Frequency bands} && \raisebox{-5pt}[0pt][0pt]{\SIrange[range-phrase=-,range-units=single]{42.5}{43.5}{\giga\hertz}} && \SIrange[range-phrase=-,range-units=single]{42.5}{43.5}{\giga\hertz} \\ 
		&                                            &&                                                                                                 && \SIrange[range-phrase=-,range-units=single]{47.2}{50.2}{\giga\hertz} \IEEEeqnarraystrutsize{0pt}{4pt}\\
		\IEEEeqnarraystrutsize{0pt}{2pt}\\
		\IEEEeqnarrayrulerow	
	\end{IEEEeqnarraybox}
\end{table}

The second simulation aims to investigate the data rate performance of the proposed multiuser \gls{mimo} downlink. The objective is to show the advantage of having multiple spatially separated antennas on the satellite in combination with \acrlong{ffr} among the user beams. The result is compared to a contemporary \gls{siso} scheme that is based on a single-reflector with \acrlong{fr4}. To highlight the importance of spatially separated antennas to benefit from a \gls{mimo} gain, the multiple-reflector approach is also compared to a \gls{mimo} downlink based on a single-antenna with multiple beams and \gls{ffr}. 

The system parameters for this simulation are provided in Table\ \ref{ParamUserLinkEval}. As already introduced in Section \ref{subsec:SystemDescription}, four reflectors form a circular array with a diameter of \SI{3}{m}, and four feeds illuminate each reflector. Each reflector has a diameter of \SI{2.6}{m}. The feed and reflector geometry is based on a satellite that is currently in orbit \cite{Fenech2016}. In our simulation scenario, the service zone of the $\AntNumSLtx=16$ spot beams is over North America, and we assume \num{250} independent and fixed single-antenna receivers per beam. The location of an individual user is randomly chosen; all users are uniformly distributed over the entire service zone. Hence, in total $\UserNumTot=16\cdot 250=\num{4000}$ households need to be served with data.\footnote{This simulation scenario and the parameter setup have partly been published in \cite{Storek2017}. For an illustration of the antenna geometry on the satellite we kindly refer to Fig.\ 1 in \cite{Storek2017}.} 

The user scheduling algorithm from \cite{Storek2017} is now applied to determine those households that form a common group. The total number of groups and the number of households in each group actually depend on several parameters. Among others, we observe a dependence on the location of the users, the threshold $\groupingThreshold$, the signal-to-noise ratio and the total number of households that have to be scheduled \cite{Storek2017}. In this simulation example, for instance, we obtained the set of $\numOfGroups=311$ user groups. 
Thus, in total $\numOfGroups=311$ \gls{mimo} downlink channel matrices $\CTMdl\in\mathds{C}^{\UserNum\times\AntNumSLtx}$ are computed. For each \gls{ut} we assume similar receiving equipment having a $\GoT$ of \SI{16.9}{dB}.

In order to show the benefit of spatial \gls{mimo} for different signal-to-noise ratios, various downlink \gls{eirp} values $\eirpdl$ are simulated in each beam ranging from \SIrange{51}{65}{dBW}. Please note again that, in the case of \gls{mimo}, $\eirpdl$ constitutes the maximum power in at least one of the \gls{mimo} beams while the power in the remaining beams can actually be equal to or lower than $\eirpdl$. This is the result of the power constraint as defined in (\ref{eq:PowerConstraintInSat}). In the \gls{siso} case instead, each beam provides the maximum downlink \gls{eirp} $\eirpdl$. Thus, the sum power over all beams of the \gls{siso} downlink is always higher or at most equals the sum power of the \gls{mimo} downlink. This ensures a fair comparison between \gls{mimo} and \gls{siso}. 

Moreover, in the \gls{ffr} scheme the entire bandwidth of \SI{500}{MHz} within the range of \SIrange{19.7}{20.2}{GHz} is available in each beam. In the case of the four-color scheme instead, only \SI{125}{MHz} per beam can be used. To ensure a fair comparison, the transmit power per beam is equal for both, the \gls{ffr} and the \gls{fr4} scheme. As a consequence, the resulting receive \gls{cnr} is even \SI{6}{\deci\bel} higher for the \gls{fr4} scheme than for the \gls{ffr} scheme. 

Since the focus is now on the performance of the multiuser downlink, the feeder uplink configuration is identical for all considered cases. In particular, a $2\times 2$ \gls{mimo} feeder link in the V-band is assumed, and the gateway antennas have a spacing of $\antspacinges=\SI{40}{\kilo\meter}$. Please note that, to support the aggregated bandwidth in the user link, the occupied bandwidth in the feeder link is adjusted accordingly. While for the \gls{fr4} scheme \SI{1}{GHz} of bandwidth is sufficient, \SI{4}{GHz} of bandwidth is necessary to support the \gls{ffr} approach. A constant uplink receive \gls{cnr} of \SI{24}{dB} is achieved in all cases.

\begin{figure}[t]
  \centering
	\includegraphics{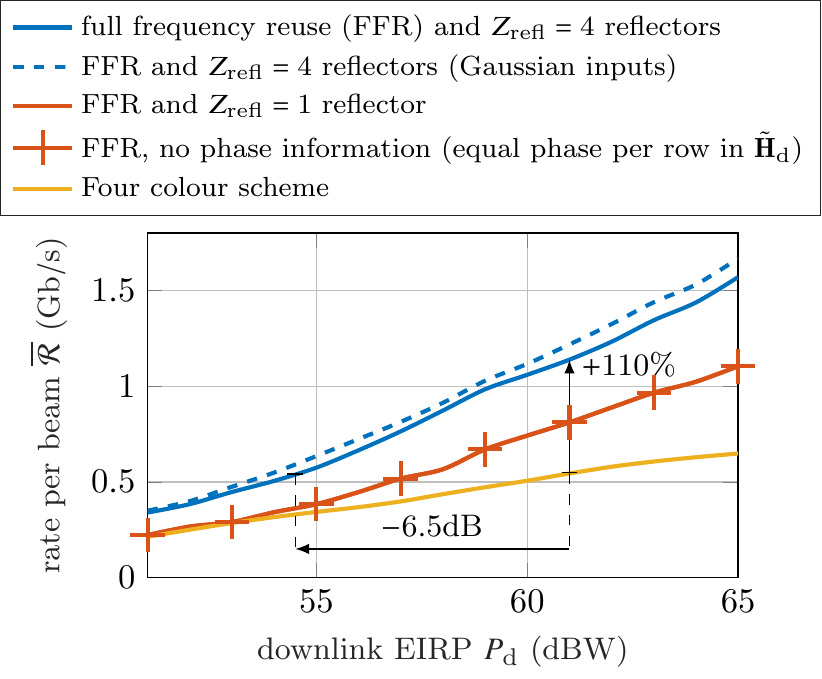}
	\caption{Achievable rate per beam as a function of the downlink \acrshort{eirp} using a \gls{mimo} feeder link with $\antspacinges=\SI{40}{\kilo\meter}$ and clear-sky conditions at both gateway stations.}
	\label{fig:sumrateperbeam}
\end{figure}

Fig.\ \ref{fig:sumrateperbeam} provides the simulation results of the sum achievable rate per user beam as a function of the downlink \gls{eirp} $\eirpdl$. The blue curve corresponds to the proposed \gls{mu-mimo} \gls{ffr} strategy with $\AntNumSLrefl=4$ reflectors. Moreover, the channel capacity based on (\ref{eq:SpeceffUser}) is shown again (blue dashed curve). The yellow curve shows the result of the state-of-the-art \gls{fr4} approach with \SI{125}{\mega\hertz} per user beam. As mentioned earlier, the result of a \gls{mimo} downlink with a single-reflector that generates all the user beams is also provided (red curve). Finally, the simulation results for the case that no phase information in the downlink channel is considered are also shown (red crosses). As mentioned in Section\ \ref{subsec:CompStateOfTheArt}, this curve represents the channel model assumptions from \cite{Joroughi2017,Christopoulos2015,Diaz2007}, in which no \added{\gls{los} channel} phase information can be exploited to form user groups or derive the precoder. \added{The absence of \gls{los} channel phase information can be manifold, for instance due to the use of a single-reflector architecture in the satellite or due to very narrow satellite and ground antenna spacings.}\deleted{The performance is actually very similar to that of using a single reflector on the satellite.} This explains why the \added{\gls{los}} phase information might not be included in the models of some state-of-the-art works. 

The \gls{mu-mimo} downlink with four separated antennas offers the highest sum achievable data rate in all considered cases. To give an example: Comparing the blue and the yellow curves at $\eirpdl=\SI{61}{dBW}$, the \gls{mu-mimo} downlink achieves \SI{1.2}{Gb/s} per beam compared to \SI{0.55}{Gb/s} only for the \gls{siso} \gls{fr4} approach. This constitutes an increase of \SI{110}{\percent}. The satellite operator has now various options how this gain can be exploited:
\begin{enumerate}
    \item Provide simply higher data rates to the households,
    \item Keep the sum data rate constant and reduce the necessary downlink \gls{eirp} instead, or
    \item exploit the gain as an additional link margin to increase the availability of the downlink.
\end{enumerate}
For example, keeping the sum achievable data rate constant would allow for a reduction of the required downlink \gls{eirp} by \SI{6.6}{dB}. This reduction could be transferred into a lower payload weight and power budget and, finally, cost reduction. As an alternative, the interference potential in terms of downlink adjacent satellite interference can be reduced.

Moreover, the \gls{mimo} downlink with four spatially separated reflectors indeed outperforms the multibeam \gls{mimo} approach based on a single-reflector. The reason is simply that the spacing of the feed elements is much smaller (in the centimeter-range) for the single-reflector approach compared to the four separated reflectors (in the meter-range). As a consequence, the spacing between two households who are part of the same group needs to be much larger\footnote{Please note again the linear relation $\antspacingsl\antspacinges\propto\distanceant{}{}\carrierwavelength$ between the antenna spacing as a function of the wavelength $\carrierwavelength$ and the \gls{ut}-satellite distance $\distanceant{}{}$, which has been introduced in Section \ref{sec:mimosatcombasics}. As a result, the smaller the spacing in orbit the larger the spacing on earth has to be.}, e.g. more than \SI{1000}{km} compared to around \SI{50}{km}. Since, in addition, the \SI{3}{dB}-contour of one beam footprint is $\approx\SI{500}{km}$, the signal energy that can be received from neighboring \gls{mimo} beams is comparably low. In other words, the interference from neighboring \gls{mimo} beams, which we are seeking to exploit as information bearing signal, is very low due to this large separation of two households who are jointly served with data. Low signal interference results in a \gls{mimo} downlink channel matrix with small values, and the multiplexing gain is ultimately limited.

To sum up, the simulation results have shown the data rate advantage of spatial multiplexing in both, the uplink and the downlink. Through multiplexing of different data streams, multiple users can be served simultaneously over the same channel. The gain of our system proposal relies on the spatial separation of the \gls{mimo} antennas.

\section{Conclusion}
\label{sec:conclusion}

In this work the theoretical basics of the \gls{mimo} \acrlong{los} concept for \acrlongpl{fss} have been presented. It has been shown that the phase of the \deleted[id=rs]{\gls{mimo} }\gls{los} signal is the key parameter to understand the possible gains of a \gls{mimo} \gls{satcom} system. By appropriately paying attention to the signal phase, the \deleted[id=rs]{requirements on the }optimal location of the \gls{mimo} antennas on earth and in orbit have been derived. 

As a promising application example, the design of future \gls{hts} systems has been addressed. Our system proposal considered both, the use of spatial multiplexing in the feeder uplink and in the multiuser downlink. While simultaneously resorting from the conventional \acrlong{fr4} to the \acrlong{ffr} scheme in the downlink, the resulting inter-beam interference can now be exploited as a useful signal energy that further increases the throughput. Based on user scheduling and \gls{mu-mimo} precoding in the gateway, single-antenna users are now able to receive individual data streams over the same channel. Simulation results have shown tremendous performance gains in terms of the sum achievable data rate in comparison to the state-of-the-art.  
\replaced[id=td]{The \gls{mimo} technology is a key enabler to boost the competitiveness of satellite communications in 5G networks.}{The \gls{mimo} technology perfectly supports the resort of conventional \gls{tv} broadcast applications to realize unicast services. Spatial multiplexing over \gls{fss} satellite systems is a promising solution to address the user's expectations to receive individual content like \gls{vod} at any time.}

\bibliographystyle{myIEEEtran}

\bibliography{library}

%

\begin{IEEEbiography}[{\includegraphics[width=1in,height=1.25in,clip,keepaspectratio]{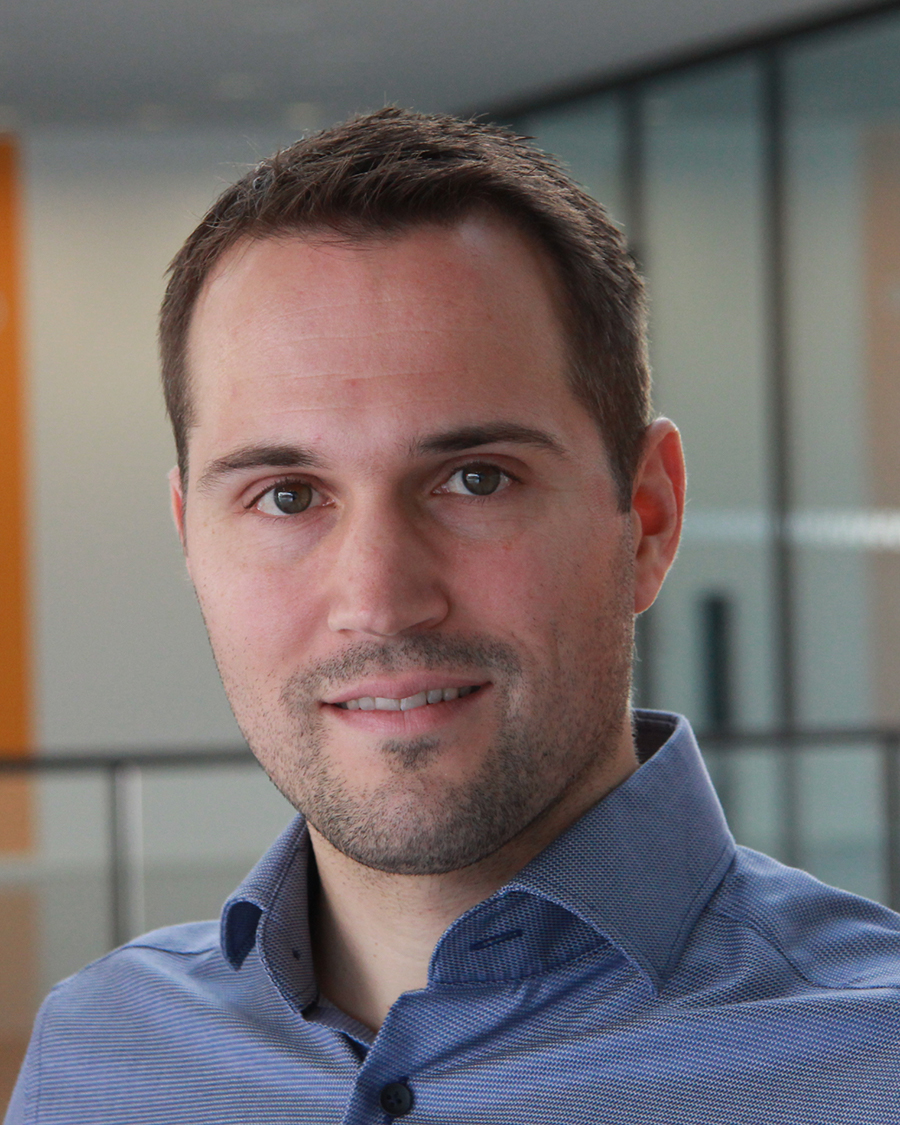}}]{Robert Schwarz}
(M’08) received the Dipl.-Ing. degree in electrical engineering and information technology from the Bundeswehr University Munich, Germany, in 2006. 

From 2006 to 2012, he was with the Federal Office of the Bundeswehr for Information Management and Information Technology, where he was involved in the German program for satellite communications of the Bundeswehr (SATCOMBw). Since 2012, he has been a Research Fellow with the Bundeswehr University Munich, Germany. His research interests include digital signal processing, waveform design, and MIMO for satellite and non-terrestrial networks. 

Mr. Schwarz is a member in the German engineers' association VDE/ITG.
\end{IEEEbiography}

\begin{IEEEbiography}[{\includegraphics[width=1in,height=1.25in,clip,keepaspectratio]{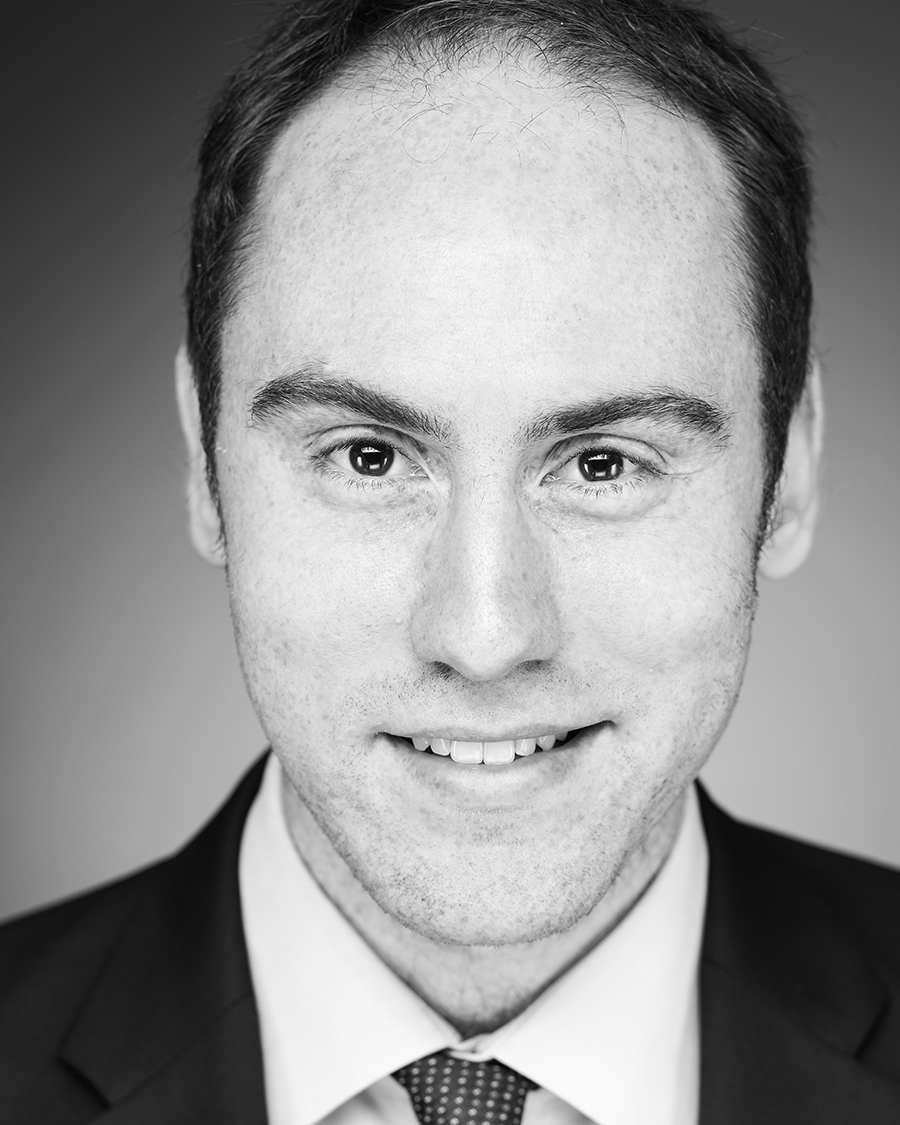}}]{Thomas Delamotte}
(M’16) was born in Reims, France, 1986. He received the master's degree in telecommunications and network engineering from the National Polytechnic Institute of Toulouse, France, in 2009. 

He is currently working towards the Ph.D. degree in electrical engineering from the Bundeswehr University Munich, Germany. Since 2010, he is working as a research fellow in the Department of Electrical Engineering and Information Technology, Bundeswehr University Munich. He leads the research group ``Digital Satellite Payloads \& Satellite Monitoring'' and is implied in several projects funded by the German Aerospace Center (DLR). His research interests include the application of advanced signal processing techniques and waveform designs for next-generation satellite systems. 
\end{IEEEbiography}

\begin{IEEEbiography}[{\includegraphics[width=1in,height=1.25in,clip,keepaspectratio]{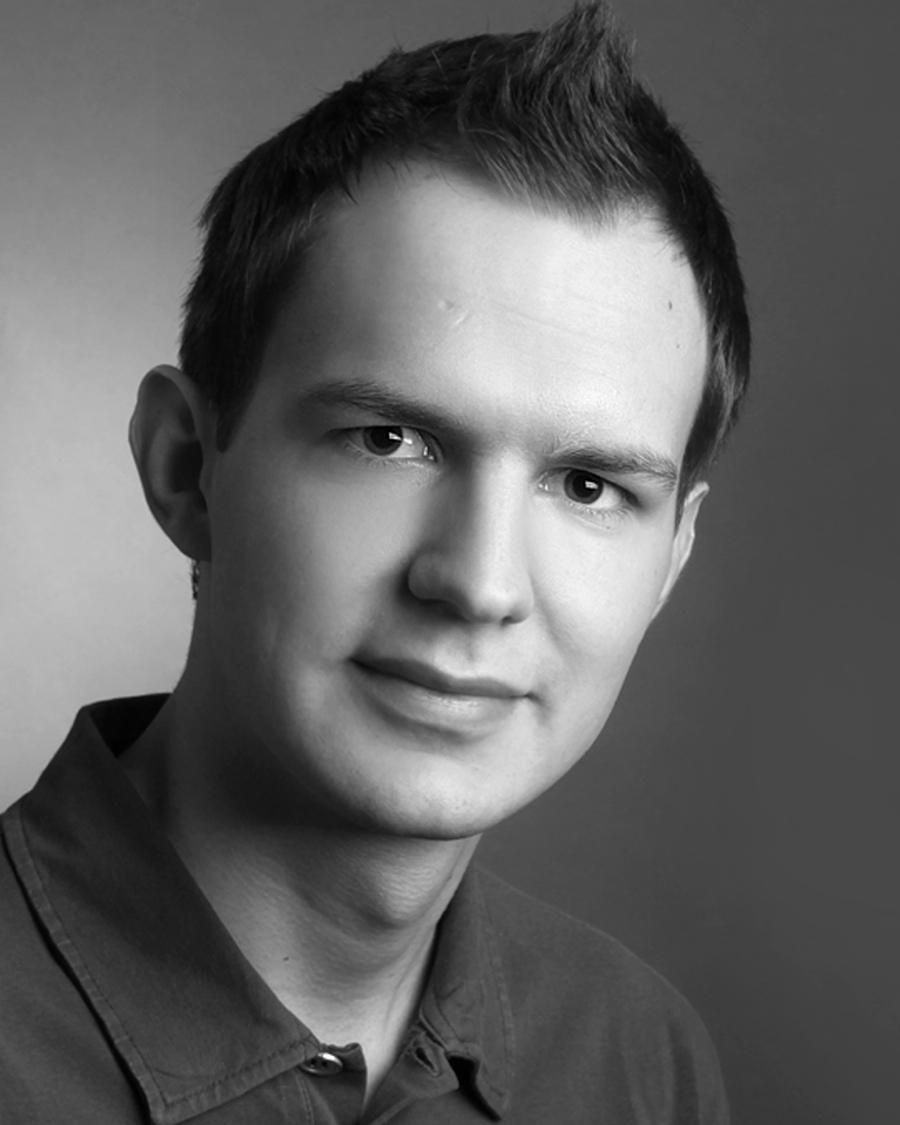}}]{Kai-Uwe Storek}
(M’15) was born in Eilenburg, Germany, in 1986. He received the Dipl.-Ing. degree in information systems engineering from the Technische Universität Dresden (TUD), in 2013. Currently he is a research associate at the Bundeswehr University Munich, Germany, working towards the Ph.D. degree.
His current research interests include MIMO signal processing, scheduling for multiuser systems and software defined radios.
\end{IEEEbiography}

\begin{IEEEbiography}[{\includegraphics[width=1in,height=1.25in,clip,keepaspectratio]{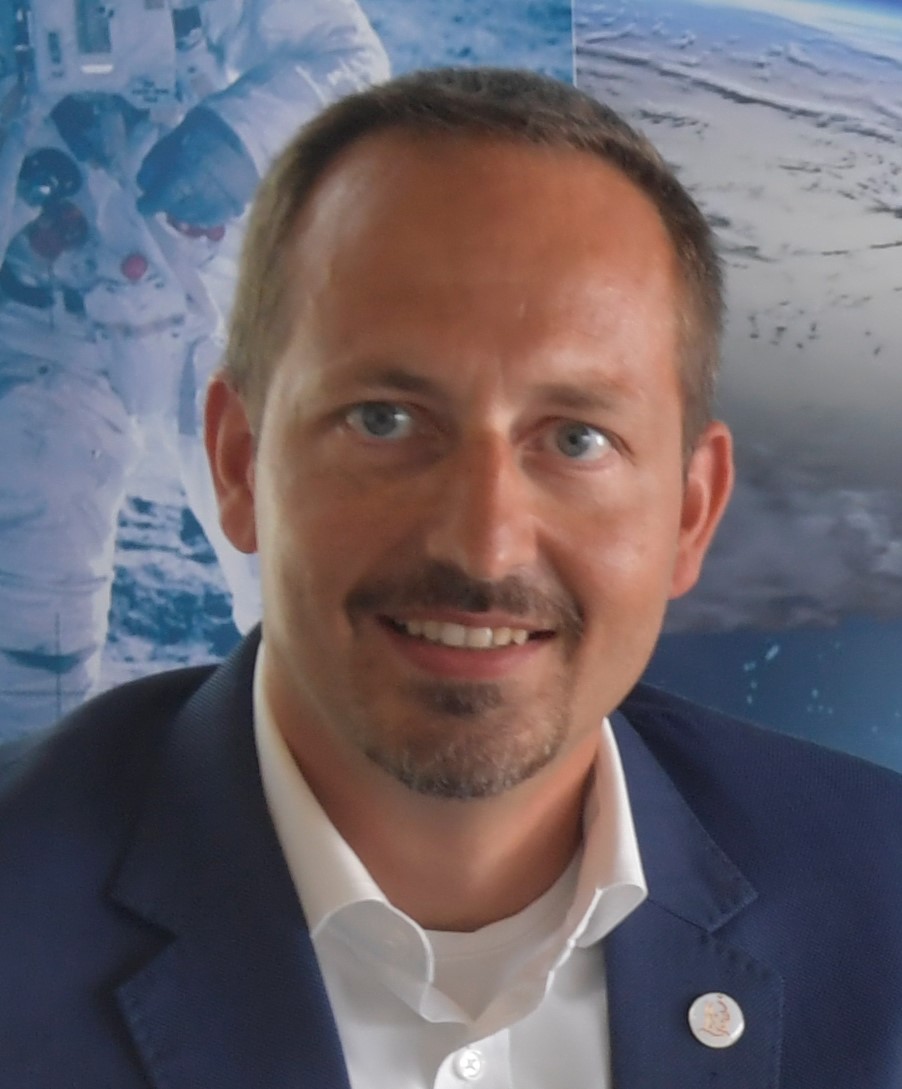}}]{Andreas Knopp}
(M’04-SM’17) earned his Ph.D. degree (with distinction) in radio communications from the Bundeswehr University Munich in 2008. 

Since 2014 he has been a Full Professor of Signal Processing, coordinating in addition Germany’s largest SpaceCom laboratory and experimental satellite ground station, the Munich Center for Space Communications. Prior to taking up the faculty position, he gained expertise as a communications engineer and satellite program manager. His current interests include satellite network integration and waveform design for 5G, digital satellite payloads, secure / antijam communications, and low power mTC. He is an entrepreneur and co-founder of two start-up companies implementing his research.

Prof. Knopp is an advisor to the German MoD, a member of the expert group on radio systems in the German engineers’ association VDE/ITG, and a member of AFCEA.
\end{IEEEbiography}

\end{document}